\documentclass[%
reprint,
twocolumn,
superscriptaddress,
 amsmath,amssymb,times, bbm,
aps,
prx,
floatfix,
]{revtex4-2}

\usepackage{libertine}
\usepackage{graphicx}
\usepackage{dcolumn}
\usepackage{bm}
\usepackage{physics}
\usepackage{bbm}
\usepackage{siunitx}
\usepackage{amsmath}
\usepackage{dsfont}
\usepackage{array,booktabs}
\usepackage{comment} 
\usepackage{xcolor}
\usepackage{rotating} 
\usepackage{textcomp, gensymb} 
\renewcommand{\arraystretch}{1.5} 
\usepackage{booktabs}
\usepackage[symbol]{footmisc}
\usepackage{placeins} 
\usepackage{float}

\definecolor{NewBlue}{rgb}{0.0313,0.090,0.208}
\definecolor{NewLightBlue}{rgb}{0,0.404,0.58}
\definecolor{NewRed}{rgb}{0.573,0.082,0.082}
\definecolor{NewBlack}{rgb}{0.,0.0,0.0}
\definecolor{NewBlue}{rgb}{0, 0, 0.41}

\usepackage[colorlinks,
    linkcolor=NewBlack,
    citecolor=NewBlue,
    urlcolor=NewBlue]{hyperref}

\newcommand{\vsi}{\textrm{V}_\textrm{Si}}
\newcommand{\vtwo}{\mathrm{V2}}
\newcommand{\gdif}{\gamma_\mathrm{d}}
\newcommand{\gion}{\gamma_\mathrm{i}}
\newcommand{\giono}{\gamma_\mathrm{i}^{0}}
\newcommand{\grecap}{\gamma_\mathrm{r}}
\newcommand{\mc}{m}
\newcommand{\Co}{C_0}
\newcommand{\Cmean}{C}
\DeclareSIUnit \vpp {\ensuremath{\mathrm{V_{pp}}}}

\begin{document}

\title{Check-probe spectroscopy of lifetime-limited emitters in bulk-grown silicon carbide}

\author{G.L. van de Stolpe}
\thanks{These authors contributed equally}
\author{L.J. Feije}
\thanks{These authors contributed equally}
\author{S.J.H. Loenen}
\thanks{These authors contributed equally}
\author{A. Das}
\author{G.M. Timmer}
\author{T.W. de Jong}
\author{T.H. Taminiau}
\email{t.h.taminiau@tudelft.nl}

\affiliation{QuTech, Delft University of Technology, PO Box 5046, 2600 GA Delft, The Netherlands}%

\affiliation{Kavli Institute of Nanoscience Delft, Delft University of Technology,
PO Box 5046, 2600 GA Delft, The Netherlands}

\date{\today}
\begin{abstract}
Solid-state single-photon emitters provide a versatile platform for exploring quantum technologies such as optically connected quantum networks. A key challenge is to ensure optical coherence and spectral stability of the emitters. Here, we introduce a high-bandwidth `check-probe' scheme to quantitatively measure (laser-induced) spectral diffusion and ionisation rates, as well as homogeneous linewidths. We demonstrate these methods on single $\vtwo$ centers in commercially available bulk-grown 4H-silicon carbide. Despite observing significant spectral diffusion under laser illumination ($\gtrsim \si{\giga \hertz \per \second}$), the optical transitions are narrow ($\sim \SI{35}{\mega \hertz}$), and remain stable in the dark ($\gtrsim\SI{1}{\second}$). Through Landau-Zener-St{\"u}ckelberg interferometry, we determine the optical coherence to be near-lifetime limited ($T_2 = \SI{16.4(4)}{\nano \second}$), hinting at the potential for using bulk-grown materials for developing quantum technologies. These results advance our understanding of spectral diffusion of quantum emitters in semiconductor materials, and may have applications for studying charge dynamics across other platforms.
\end{abstract}


\maketitle


Optically active solid-state defects have enabled pioneering experiments in the field of distributed quantum computation 
\cite{waldherr_quantum_2014,cramer_repeated_2016,abobeih_faulttolerant_2022, debone_thresholds_2024a} and quantum networks \cite{pompili_experimental_2022,knaut_entanglement_2024}. Proof-of-principle experiments have demonstrated primitives for quantum error correction \cite{waldherr_quantum_2014,cramer_repeated_2016,abobeih_faulttolerant_2022} and the realisation of a three-node network \cite{pompili_realization_2021, hermans_qubit_2022}. Key to these applications is the ability to connect multiple emitters via their coherent spin-optical interface, with many applications requiring narrow, stable optical transitions \cite{pompili_realization_2021,knaut_entanglement_2024}.

Spectral diffusion of the transitions, caused by fluctuating charge impurities within the bulk material or at the surface, poses a major challenge, especially when emitters are integrated in nanostructures \cite{wolters_measurement_2013, orphal-kobin_optically_2023, faraon_coupling_2012,ruf_optically_2019}. Moreover, laser pulses used to probe or manipulate the emitter can exacerbate such diffusion \cite{orphal-kobin_optically_2023, robledo_control_2010}. Experimental techniques that enable the quantitative study of spectral diffusion and its timescales provide insight into the environmental charge dynamics, potentially allow for targeted optimisation of material properties and fabrication processes, and enable pathways to mitigate diffusion through pre-selection \cite{robledo_control_2010, bernien_heralded_2013}. However, commonly used methods may significantly disturb the system through continuous laser illumination, complicating the unambiguous determination of transition linewidths and diffusion rates under different laser illumination conditions \cite{heiler_spectral_2024,koch_limits_2023,orphal-kobin_optically_2023, wolters_measurement_2013, candido_suppression_2021}.

Here, we introduce a comprehensive check-probe spectroscopy toolbox for characterising and mitigating spectral diffusion of single solid-state emitters. Our methods offer high-bandwidth, quantitatively extract diffusion and ionisation rates, and introduce minimal system disturbance from laser illumination, enabling accurate measurements even in heavily diffusive environments. Additionally, our work provides a framework for the quantitative analysis of heralded preparation of the charge environment, which has become an indispensable tool to mitigate spectral diffusion in quantum network and other experiments \cite{robledo_control_2010,bernien_heralded_2013, brevoord_heralded_2024, pompili_realization_2021, hermans_qubit_2022, abobeih_faulttolerant_2022}.

 
We apply these methods to study single k-site $\vsi$ ($\vtwo$) centers (a next-generation candidate for quantum networks \cite{babin_fabrication_2022,lukin_integrated_2020,heiler_spectral_2024}), embedded in nanopillars etched in commercially available bulk-grown 4H-silicon carbide (SiC) \cite{hausmann_fabrication_2010,radulaski_scalable_2017}. This system exhibits a high degree of spectral diffusion ($>\SI{1}{\giga \hertz}$ diffusion-averaged linewidth), typical for single quantum emitters in bulk-grown silicon or silicon carbide \cite{macquarrie_generating_2021, anderson_electrical_2019}. First, we determine spectral diffusion rates with and without laser illumination. Using this knowledge, we select configurations of the system with narrow spectral transitions, which can be tuned over the breadth of the inhomogeous linewidth and can be stored for over a second and accessed on-demand. Finally, through the observation of Landau-Zener-St{\"u}ckelberg interference \cite{shevchenko_landau_2010}, we determine the optical coherence time to be: $T_2 = \SI{16.4(4)}{\nano \second}$, consistent with the lifetime limit for these defects \cite{liu_silicon_2024}. 

Although high-purity epitaxial layers provide a starting point with less spectral diffusion \cite{babin_fabrication_2022,heiler_spectral_2024,fang_experimental_2024} (\ref{sec:PLE_literature_review}), our observation of lifetime-limited coherence in nanostructures in bulk-grown silicon carbide, hints towards the possibility of using such mass-fabricated material for quantum technology development and applications. Furthermore, the techniques developed here might facilitate the targeted optimisation of material and fabrication recipes, and can be readily transferred to other platforms \cite{anderson_electrical_2019, higginbottom_optical_2022, brevoord_heralded_2024, neuhauser_correlation_2000,iff_substrate_2017}. 

\begin{figure*}[ht]
  \includegraphics[width=1 \textwidth]{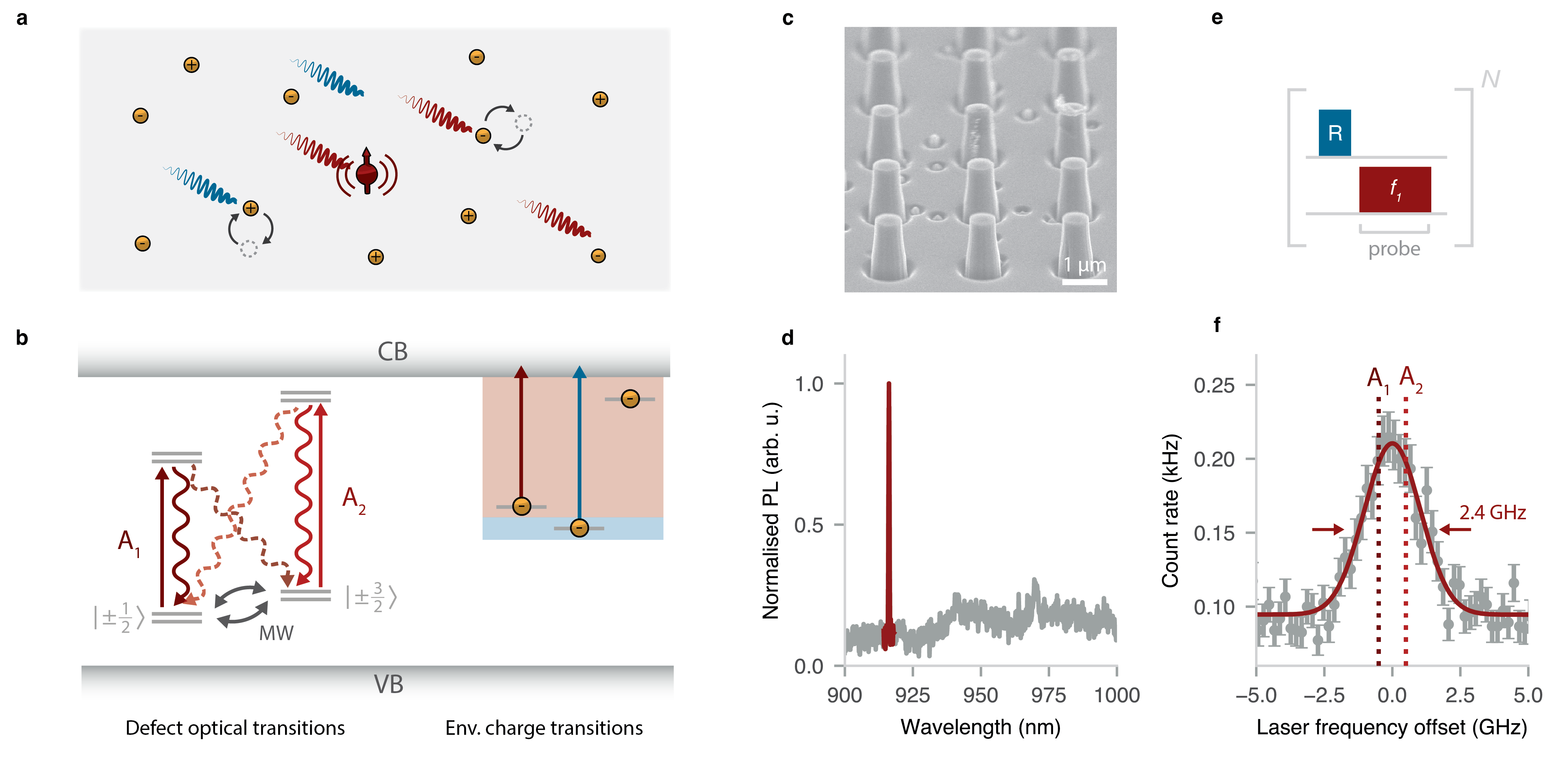}
  \caption{\label{fig:intro} \textbf{Emitter optical properties and laser-induced charge dynamics.} \textbf{a)} Schematic of the system. A single $\vtwo$ center in SiC is surrounded by charges (yellow circles) associated to intrinsic residual impurities \cite{son_charge_2021}. Under laser illumination, these charges can be mobilised after excitation to (from) the conduction (valence) band, indicated by blue and red wiggly lines. \textbf{b)} Energy diagram, depicting the $\vtwo$ center's optical transitions (left) and possible laser-induced charge dynamics of the (unknown) impurities in the environment (right). The spin-dependent $\mathrm{A_1}$ and $\mathrm{A_2}$ transitions can be excited with a tunable, near-infrared (NIR) laser (916 nm, red arrow), while a high-energy repump laser (785 nm, blue arrow) is used to scramble the charge state of the $\vtwo$ center and its environment. The ground-state spin ($S=\frac{3}{2}$) can be manipulated with microwave (MW) radiation. \textbf{c)} Scanning-electron-microscopy image of a sample used in this work, which is diced from a $\SI{4}{inch}$ commercially available 4H-SiC bulk wafer. Nanopillars ($\sim \SI{500}{\nano \meter}$ diameter) are fabricated to improve the photon collection efficiency. \textbf{d)} Representative low-temperature (4K) emission spectrum of a $\vtwo$ center under repump-laser excitation, showing the characteristic zero-phonon line at \SI{916}{\nano \meter} (red highlight). \textbf{e)} Experimental sequence of the diffusion-averaged photoluminescence-excitation spectroscopy (PLE). The frequency $f_1$ of the NIR laser (red) is scanned over the $\vtwo$ zero-phonon line, while emission in the phonon sideband is collected. The repump laser (blue) scrambles the charge state of the emitter and its environment before every repetition (total $N$). \textbf{f)} Measured PLE spectrum. Averaging over many charge-environment configurations results in a single, broad peak ($\SI{2.4(1)}{\giga \hertz}$ FWHM) that encompasses the $A_1$ and $A_2$ transitions (separated by $\sim \SI{1}{\giga \hertz}$). The laser frequency is offset from $\SI{327.10}{\tera \hertz}$.
  }
\end{figure*}

\section*{Results}
\subsection*{System: single V2 centers in nano-structured bulk-grown silicon carbide}

We consider spectral diffusion caused by fluctuating charges in the environment of the emitter, for example associated to material impurities or surface defects that modify the optical transition frequency via the Stark shift \cite{ruhl_stark_2020,lukin_spectrally_2020, pieplow_quantum_2024,ji_correlated_2024}. Although these dynamics are largely frozen at cryogenic temperatures \cite{candido_suppression_2021}, charges can still be mobilised through laser illumination used for the optical addressing of the emitter \cite{anderson_electrical_2019,candido_suppression_2021,orphal-kobin_optically_2023, wolters_measurement_2013} (Fig. \ref{fig:intro}a). In particular, charges can be excited to the conduction (or valence) band via a single-photon process if the energy difference from the occupied charge state is smaller than the associated energy of the laser frequency (Fig.\ref{fig:intro}b). Subsequent decay to a different spatial position causes fluctuations in the electric field at the location of the emitter \cite{orphal-kobin_optically_2023,candido_suppression_2021}. 

This work considers single k-site $\vsi$ ($\vtwo$) centers in commercially available bulk-grown silicon carbide at 4K. In this material, diffusion is likely caused by charges associated with residual defects and shallow dopants (concentrations $\sim10^{15}\SI{}{\per\centi\meter\cubed}$) that are created during the growth process \cite{son_charge_2021}. We apply two types of lasers: a 785 nm `repump' laser for charge-state reinitialisation ($\sim\SI{10}{\micro\watt}$), and two frequency-tunable near-infrared (`NIR') lasers ($\sim\SI{10}{\nano\watt}$) for resonant excitation of the $\vtwo$ center's spin-dependent $A_1$ and $A_2$ zero-phonon-line transitions (Fig. \ref{fig:intro}b and d) \cite{babin_fabrication_2022}.  

We fabricate nanopillars that enhance the optical collection efficiency (Methods), to mitigate the effects of the unfavourable dipole orientation in c-plane 4H-SiC (Fig. \ref{fig:intro}c). In about one in every 10 pillars, we observe a low-temperature spectrum with a characteristic zero-phonon line at $\SI{916}{\nano\meter}$ (Fig. \ref{fig:intro}d), hinting at the presence of single $\vtwo$ centers confined to the nanopillars. The dimensions of the nanopillar, with a diameter of $\sim \SI{500}{\nano\meter}$ and a height of $\sim \SI{1.2}{\micro\meter}$, mean that surface- and fabrication-related effects might contribute to the diffusion dynamics.

\subsection*{Photoluminescence excitation spectroscopy}
First, we measure the $\vtwo$ diffusion-averaged optical absorption linewidth via photoluminescence excitation spectroscopy (PLE). By repeatedly interleaving repump pulses (`R', $\SI{10}{\micro\second}$, $\SI{10}{\micro\watt}$) with NIR pulses at a varying frequency $f_1$ ($\SI{10}{\micro\second}$, $\SI{10}{\nano\watt}$, Fig. \ref{fig:intro}e), we randomise the $\vtwo$ charge environment before each repetition, effectively averaging over many spectral configurations. In a system without spectral diffusion, we would expect to observe two distinct narrow lines (FHWM of $\sim \SI{26}{\mega\hertz}$ and $\sim \SI{11}{\mega\hertz}$), separated by $\Delta \approx \SI{1}{\giga\hertz}$ \cite{udvarhelyi_vibronic_2020,babin_fabrication_2022, liu_silicon_2024}, corresponding to the separation of the $\mathrm{A_1}$ and $\mathrm{A_2}$ transitions. However, we observe a broad Gaussian peak ($\SI{2.4(1)}{\giga\hertz}$, see Fig. \ref{fig:intro}f), hinting at a high degree of spectral diffusion, consistent with comparable experiments in similar bulk semiconductor materials \cite{anderson_electrical_2019,higginbottom_optical_2022}.

\begin{figure}[ht]
  \includegraphics[width=1 \columnwidth]{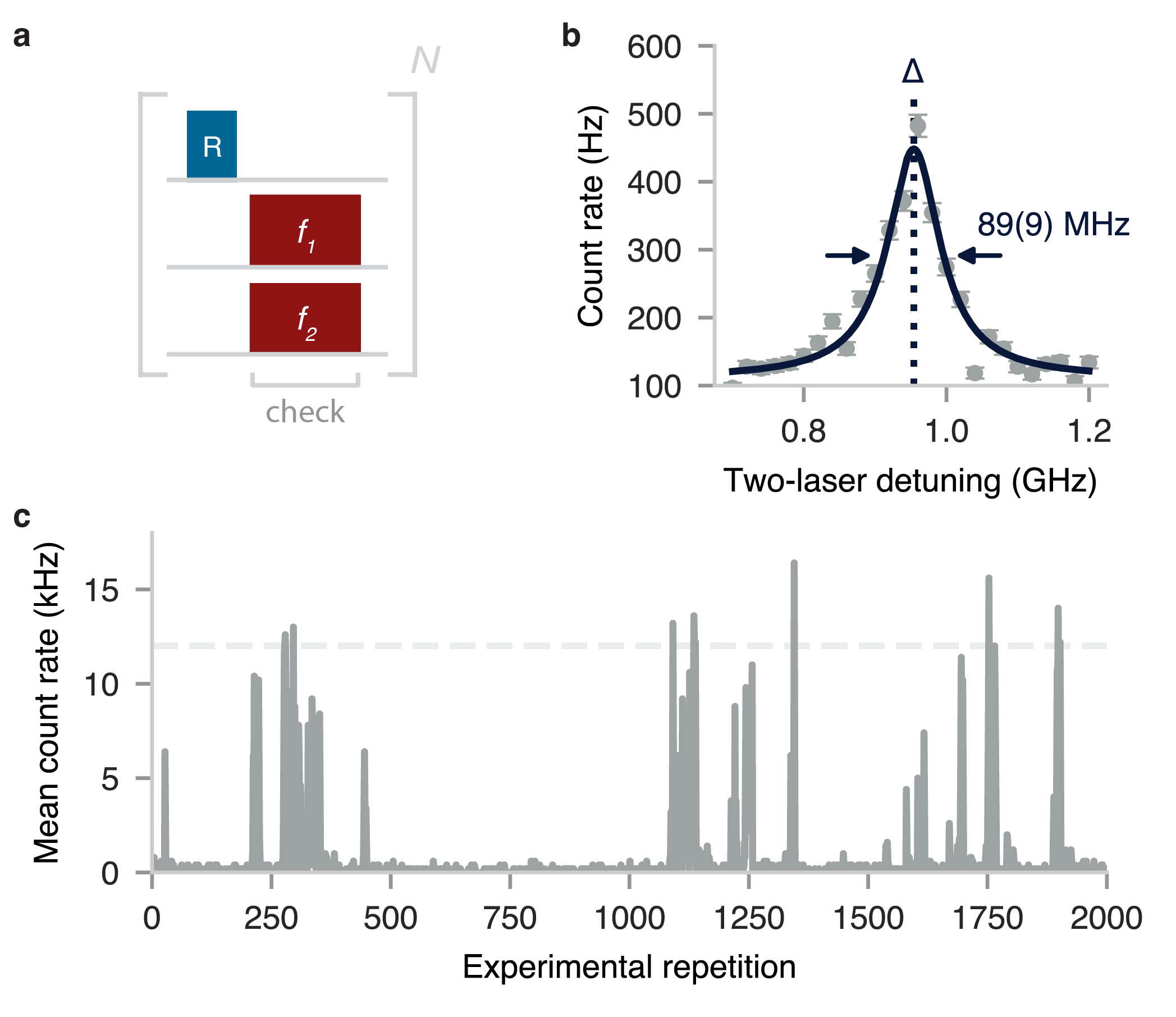}
  \caption{\label{fig:resonance_check} \textbf{Two-laser photoluminescence excitation spectroscopy.} \textbf{a)} Experimental sequence. A short (\SI{10}{\micro \second}, \SI{10}{\micro \watt}) high-energy (785 nm) repump laser pulse (`R') partially scrambles the charge state of both the environment and the $\vtwo$ center. Emission is collected during the `check' block, when two NIR lasers at frequencies $f_1$ and $f_2$ are turned on (approximately resonant with the broad peak in Fig. \ref{fig:intro}g). \textbf{b)} We observe an increase in count rate if the laser frequency difference is equal to $ \Delta = \SI{954(2)}{\mega \hertz}$, the spacing between the $\mathrm{A_1}$ and $\mathrm{A_2}$ transitions. A Lorentzian fit obtains a FWHM of $\SI{89(9)}{\mega\hertz}$. \textbf{c)} Detected mean count rate per experimental repetition, when the length of the `check' block is set to \SI{5}{\milli \second}. In most repetitions, the defect is off-resonant with the lasers. When the $\mathrm{A_1}$ and $\mathrm{A_2}$ transitions coincide with laser frequencies $f_1$ and $f_2$, we observe significant emission ($\gg\SI{1}{\kilo \hertz}$). Thresholding (dashed line) on the detected counts can be employed to prepare specific (i.e. `on resonance') spectral configurations of the $\vtwo$ center.
  }
\end{figure}

In order to probe the individual $\mathrm{A_1}$ and $\mathrm{A_2}$ transitions, we employ a two-laser PLE scan \cite{higginbottom_optical_2022}. Compared to the sequence in figure \ref{fig:intro}e, we now fix frequency $f_1$ close to the middle of the broad resonance (Fig. \ref{fig:intro}f) and add a second NIR laser at frequency $f_2$ (Fig. \ref{fig:resonance_check}a). We observe a significant increase in the detected count rate when the frequency difference satisfies: $f_2 - f_1 \approx \Delta = \SI{954(2)}{\mega \hertz}$, explained by a strong reduction in optical pumping (which otherwise quickly diminishes the signal \cite{higginbottom_optical_2022}). Importantly, the relatively narrow resonance condition (FWHM of $\SI{89(9)}{\mega \hertz}$) observed in Fig. \ref{fig:resonance_check}b suggests that the homogeneous linewidth is much narrower than the diffusion-averaged linewidth in Fig. \ref{fig:intro}f. 

Next, we fix the laser frequency difference to $\Delta$ and record the counts per experimental repetition (Fig. \ref{fig:resonance_check}c). We obtain a telegraph-like signal, consistent with a single $\vtwo$ center that is spectrally diffusing. Such a signal allows for the implementation of a charge-resonance check \cite{bernien_heralded_2013, brevoord_heralded_2024}, which probes whether the $\vtwo$ center is in the desired negative charge state, and its two transitions are resonant with the NIR lasers. If the number of detected counts passes a threshold $T$ (e.g. the grey dashed line in Fig. \ref{fig:resonance_check}c), we conclude that the defect was on resonance in that specific experimental repetition, allowing for post-selection (or pre-selection) of the data. In the following, we will explore how such post-selection tactics can be exploited to gain insights in the spectral diffusion dynamics.

\subsection*{Check-probe spectroscopy: ionisation and spectral diffusion dynamics}
Next, we develop a method to measure the ionisation and spectral diffusion dynamics of the $\vtwo$ center. Currently, various experimental techniques exist, based either on tracking the transition frequency with subsequent PLE scans \cite{orphal-kobin_optically_2023,koch_limits_2023,heiler_spectral_2024}, or on autocorrelation-type measurements \cite{wolters_measurement_2013,fleury_spectral_1993}. The former method struggles with measuring dynamics faster than the acquisition timescale of a single scan \cite{basché_singlemolecule_1997, heiler_spectral_2024} (see also Supplementary Fig. \ref{fig:SI_Fast_PLE_plot}). The latter, although fast, offers limited flexibility for probing diffusion under external perturbations \cite{wolters_measurement_2013}. 

\begin{figure}[ht]
  \includegraphics[width=1 \columnwidth]{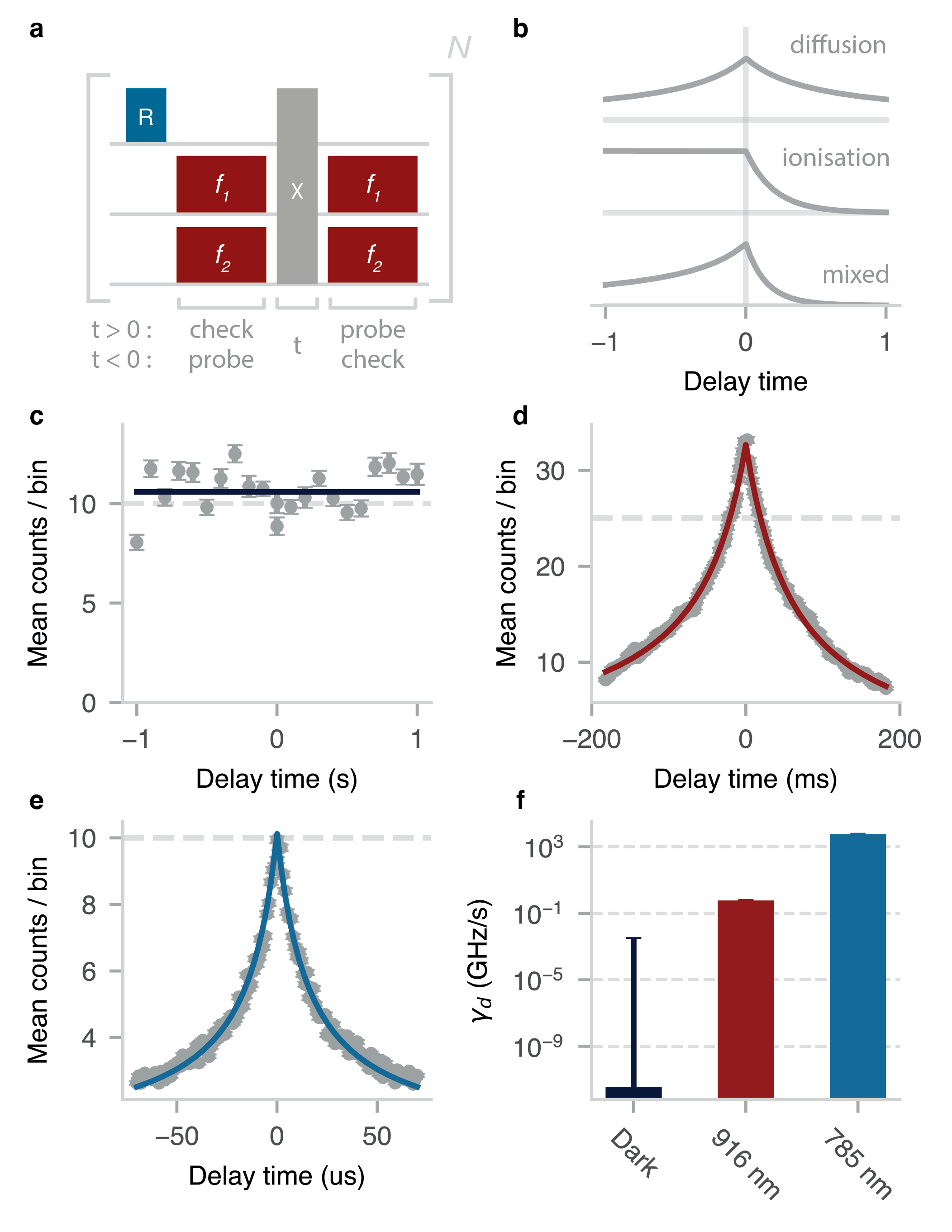}
  \caption{\label{fig:spectral_diffusion_dynamics} \textbf{Spectral diffusion dynamics.} \textbf{a)} Experimental sequence. A `check' block (\SI{2}{\milli \second}, \SI{20}{\nano \watt}) is followed by a system perturbation (marked `X'), which here consists either of turning off the lasers (c), turning on the NIR lasers (d), or turning on the repump laser (e). A second block (\SI{2}{\milli \second}, \SI{20}{\nano \watt}) probes whether the defect has diffused away, or has ionised (denoted `probe'). Data is post-selected by imposing a minimum-counts threshold ($T$), heralding the emitter on resonance
  in the first (second) block and computing the mean number of counts in the second (first) block, which encodes the emitter brightness at future (past) delay times $t$. \textbf{b)} Schematic illustrating the expected signal (according to Eq. \ref{eq:mean_counts_dynamics}), when either ionisation or spectral diffusion is dominant (setting $\grecap \approx 0$). \textbf{c)} No significant spectral diffusion or ionisation is observed when the lasers are turned off. The solid line is a fit to the data using Eq. \ref{eq:mean_counts_dynamics}. Dashed grey line denotes the set threshold (in a \SI{2}{\milli \second} window). \textbf{d)} Experiment and fit under \SI{20}{\nano \watt} of NIR laser power (916 nm). \textbf{e)} Experiment and fit under \SI{1}{\micro \watt} of repump laser power. \textbf{f)} Extracted saturation-diffusion rates, obtained at laser powers of $\sim \SI{20}{\nano \watt}$ (resonant) and $\sim \SI{5}{\micro \watt}$ (repump). See supplementary Fig. \ref{fig:SI_laser_power_sweep} for underlying data and error analysis.
  }
\end{figure}

Here, we take a different approach, based on a pulsed \emph{check-probe} scheme as outlined in Fig. \ref{fig:spectral_diffusion_dynamics}a. Following a charge-randomisation (repump) step, two `check' blocks are executed (as in Fig.\ref{fig:resonance_check}a), separated by a perturbation of the system (grey block marked `X'). Such a perturbation might consist of turning on (or off) specific lasers (e.g. NIR or repump) during the delay time $t$. This pulsed scheme features both high bandwidth (limited by the operating speed of the lasers), and allows one to isolate diffusion originating from the perturbation from other sources. 

Importantly, one can either post-select on high counts (i.e. `check') in the block before, or after the perturbation, effectively initialising the emitter on resonance at the start, or at the end of the experiment. By `probing' the emitter brightness after (before) the perturbation, we effectively track its evolution forward (backward) in time, denoted as delay time $t>0$ ($t<0$) in Fig. \ref{fig:spectral_diffusion_dynamics}a. This allows for the distinction between time-symmetric and non-time-symmetric perturbation processes (e.g. spectral diffusion or ionisation of the emitter, see Fig. \ref{fig:spectral_diffusion_dynamics}b), as opposed to evaluating the purely symmetric autocorrelation function \cite{wolters_measurement_2013}.

To quantitatively describe the signal, we derive an analytical expression that takes into account spectral diffusion and ionisation of the emitter. In this system, spectral diffusion is mainly caused by laser-induced reorientation of charges surrounding the defect, whose dynamics can be approximated by a bath of fluctuating electric dipoles \cite{candido_suppression_2021}. To model this, we employ the spectral propagator formalism \cite{zumofen_spectral_1994,basché_singlemolecule_1997}, which describes the evolution of the spectral probability density function in time, and whose form is given by a Lorentzian with linearly increasing linewidth $\gamma(t) = \gamma_d \,\abs{t}$ \cite{zumofen_spectral_1994} (with $\gdif$ the effective diffusion rate). Note that this description is valid only at short timescales ($\gamma_d  \,\abs{t} \ll \SI{1}{\giga\hertz}$), as the spectral probability density should eventually converge to the diffusion-averaged distribution observed in Fig. \ref{fig:intro}f \cite{zumofen_spectral_1994}. Furthermore, we take the spectral propagator to be time-symmetric, and model the 
ionisation (charge recapture) of the emitter as an exponential decay of fluorescence, governed by rate $\gion$ ($\grecap$). 

The mean number of observed counts at delay time $t$ can be described by (\ref{sec:model_derivation}):
\begin{equation} \label{eq:mean_counts_dynamics}
  \Cmean(t) = \begin{cases}
     \Co \left( 1 + \gdif \, t\, /\,\Gamma \right)^{-1} \,e^{-\gion t} , & \text{if $t > 0$}.\\
     \Co \left( 1 - \gdif \, t\, /\,\Gamma \right)^{-1} \,e^{\grecap t}, & \text{otherwise}.
  \end{cases}
\end{equation}
with $\Co$ the mean number of observed counts at $t=0$, $\Gamma$ the emitter's (Lorentzian) homogeneous linewidth, and $\gdif, \gion, \grecap > 0$. Note that Eq. \ref{eq:mean_counts_dynamics} in general does not obey time-inversion symmetry (for $\gion \neq \grecap$), and in specific cases allows for a clear distinction between ionisation and diffusion processes (e.g. if $\grecap \approx 0$, see Fig. \ref{fig:spectral_diffusion_dynamics}b). Next to that, the functional form of Eq. \ref{eq:mean_counts_dynamics} captures information about the type of processes at play: emitter charge dynamics are described by an exponential decay while spectral diffusion has a power law dependence.

We experimentally implement the sequence for three distinct perturbations: (i) no laser illumination (Fig. \ref{fig:spectral_diffusion_dynamics}c), (ii) illumination with the two NIR lasers (\SI{20}{\nano \watt}, Fig. \ref{fig:spectral_diffusion_dynamics}d), and (iii) illumination with the repump laser (\SI{1}{\micro \watt}, Fig. \ref{fig:spectral_diffusion_dynamics}e). We observe a wide range of dynamics, from the microsecond to second timescale, and observe excellent agreement between the data and the model (solid lines are fits to Eq. \ref{eq:mean_counts_dynamics}). 

To quantitatively extract ionisation and diffusion rates under the perturbations, we set $\Gamma = \SI{36}{\mega \hertz}$, (independently determined in Fig. \ref{fig:lifetime_limited}e). We find that extracted rates are weakly dependent on the set threshold value, resulting from non-perfect initialisation on-resonance, but converge for higher $T$ (\ref{sec:threshold_dependence_rates}). Averaging over a range of threshold values, we find diffusion rates $\gdif= 0.00(2)\,\si{\giga \hertz \per \second}, 0.60(2)\,\si{\giga \hertz \per \second} , \num{2.4(2)e3}$ \si{\giga \hertz \per \second}, for perturbations (i), (ii) and (iii), respectively. In the dark, where almost no diffusion is apparent, the fit only converges if we set $\gion, \grecap = 0$, which is a reasonable assumption at \SI{4}{\kelvin}, given the deep-level nature of the $\vtwo$ center \cite{candido_suppression_2021, son_charge_2021}. 

Ionisation effects are only observed under the NIR-laser perturbation, due to the short diffusion timescale during the repump-laser perturbation (resulting in divergent fit results for $\gion$ and $\grecap$). From the data in Fig. \ref{fig:spectral_diffusion_dynamics}d, we extract $\gion = 1.0(2)$ \si{\hertz} and $\grecap = 0.03(4)$ \si{\hertz}. Correcting for reduced ionisation when the $\vtwo$ center is off-resonance with the NIR lasers results in an ionisation rate $\giono = \SI{3(1)}{\hertz}$ (see \ref{sec:model_derivation}). 

We repeat the experiments for various laser powers, and observe a saturation-type behavior of the diffusion rates, both under NIR-lasers and repump laser excitation (Supplementary Fig. \ref{fig:SI_laser_power_sweep}). The higher saturation-diffusion rate measured for the repump laser is likely due to the larger fraction of charge traps that can be ionised via a single-photon process (see Fig.\ref{fig:intro}b and Fig. \ref{fig:spectral_diffusion_dynamics}f). Different $\vtwo$ centers in the material, show some variation in the (saturation) diffusion rates (Supplementary Figs. \ref{fig:SI_Fast_PLE_plot} and \ref{fig:SI_laser_power_sweep_2}. Note that the behaviour at powers beyond those accessed in these experiments will determine if spectral stability persists under repeated fast optical $\pi$-pulses, as commonly used for remote entanglement generation experiments \cite{brevoord_heralded_2024,pompili_realization_2021, bernien_heralded_2013}.

\subsection*{Check-probe spectroscopy: linewidth} \label{sec:superfast_ple}
\begin{figure}[ht]
  \includegraphics[width=1 \columnwidth]{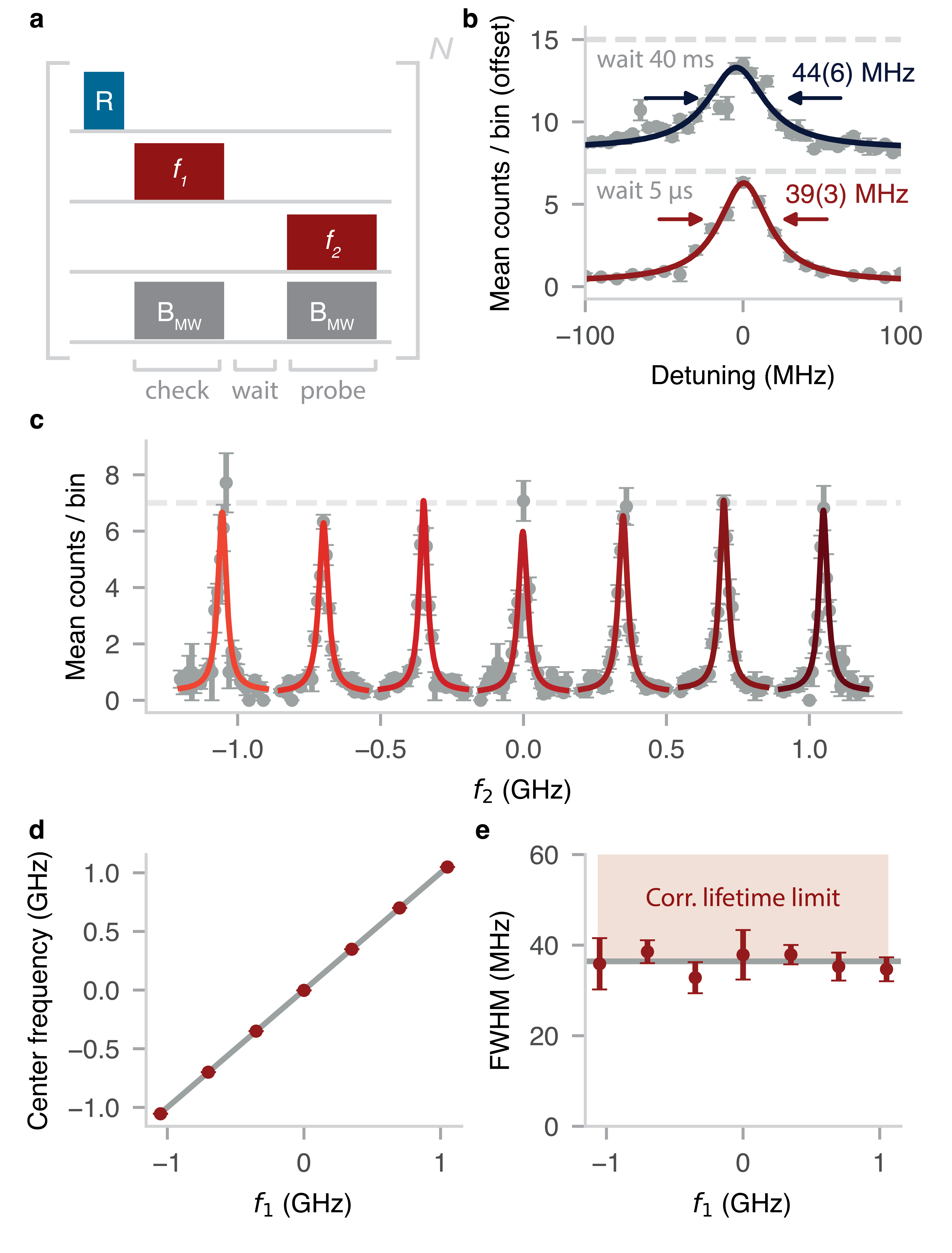}
  \caption{\label{fig:lifetime_limited} \textbf{Homogeneous linewidth and state preparation} \textbf{a)} Experimental sequence. A resonant laser at $f_1$ and a microwave (MW) pulse resonant with the ground-state spin transition act as a resonance check, initialising the optical transition near $f_1$. Next, a second laser probes the defect emission at $f_2$ (for \SI{2}{\milli \second}), yielding a measurement of the inewidth with minimal disturbance. \textbf{b)} Experimental data showing narrow optical transitions, with the bottom (top) data corresponding to a waiting time of \SI{5}{\micro \second} (\SI{40}{\milli \second}) between the $f_1$ and $f_2$ laser pulses (offset for clarity). Data are fitted to a Lorentzian with a FWHM of \SI{39(3)}{\mega \hertz} (\SI{44(6)}{\mega \hertz}). The counts threshold is set to $T = 7$ during the $f_1$ pulse. \textbf{c)} Data and fits as in (b)(bottom), scanning $f_2$ around $f_1$, when $f_1$ is set at different frequencies (various shades of red) within the broad diffusion-averaged line measured in Fig. \ref{fig:intro}g. \textbf{d)} Measured resonance center frequency as a function of the set $f_1$ frequency (solid grey line: $f = f_1$). The defect emission frequency can effectively be tuned over a \si{\giga \hertz} range. \textbf{e)} Corresponding linewidths extracted from (c), with inverse-variance weighted mean \SI{36(1)}{\mega \hertz} (solid grey line). The shaded region denotes the expected minimum linewidth ($\sim \SI{36}{\mega \hertz}$), given the lifetime limit of $\sim \SI{20}{\mega \hertz}$, and correcting for power broadening ($\sim \SI{26}{\mega \hertz}$, Fig. \ref{fig:LZS_interference}j) and residual inhomogeneous broadening ($\sim \SI{15}{\mega \hertz}$, Eq. \ref{eq:spectral_probability_density}).
  }
\end{figure}

Having established the spectral diffusion timescales, we characterise the homogeneous linewidth with minimal laser-induced disturbance. We use an optical spectroscopy method based on the check-probe scheme (similar to Ref. \cite{hermans_entangling_2023}) that, in this system, only requires laser illumination on timescales short compared to the laser-induced diffusion timescales (contrary to commonly used `scanning' PLE \cite{koch_limits_2023,heiler_spectral_2024,orphal-kobin_optically_2023, babin_fabrication_2022, lukin_twoemitter_2023, robledo_control_2010}, Supplementary Fig. \ref{fig:SI_Fast_PLE_plot}). 

First, we execute an alternative implementation of the `check' block (compared to Figs \ref{fig:resonance_check}a and \ref{fig:spectral_diffusion_dynamics}a), that consists of a single NIR-laser pulse at $f_1$, together with a MW pulse that mixes the spin states (Fig. \ref{fig:lifetime_limited}a). By post-selecting on high counts, either the $\mathrm{A_1}$ or $\mathrm{A_2}$ transition is initialised on-resonance with $f_1$. A second laser is used to probe the defect emission at a frequency $f_2$ immediately thereafter ($\sim \si{\micro \second}$ timescale, here limited by the microprocessor clock cycle). By studying the mean number of counts during the $f_2$ pulse, (an upper bound for) the homogeneous linewidth can be extracted (Fig. \ref{fig:lifetime_limited}b).

We introduce a quantitative model for the signal that extracts the homogeneous linewidth and considers the residual inhomogeneous broadening resulting from non-perfect initialisation on-resonance. To this end, we compute the spectral probability density immediately after the `check' block, as a function of the number of detected photons $\mc \geq T$ using Bayesian inference (see \ref{sec:bayesian_signal_analysis}):
\begin{equation} \label{eq:spectral_probability_density}
    P(f\,|\,\mc \geq T) = \frac{1}{N_T} \left(1 - \Gamma_\mathrm{i}\left[  T, \lambda(f-f_1) \right]\right) \,,
\end{equation}
with $f$ the emitter frequency, $\lambda(f)$ the pure (i.e. homogeneous) spectral response of the emitter, $\Gamma_\mathrm{i}[a,z]$ the incomplete Gamma function and $N_T$ a normalisation constant (\ref{sec:bayesian_signal_analysis}). The expression in Eq. \ref{eq:spectral_probability_density} is strongly dependent on $T$, with higher threshold values leading to distributions that are sharply peaked around $f_1$. Note that this analysis assumes negligible laser-induced diffusion during the `check' block, placing limits on the used laser power and the block's duration ($\ll 1/\gdif$).

The measured signal, i.e. the mean number of detected counts in the `probe' block is then given by:
\begin{equation} \label{eq:fast_spectroscopy_signal}
    \Cmean(f) = P(f\,|\,\mc \geq T) \, * \, \lambda(f)  \, ,
\end{equation}
where $*$ denotes the linear convolution. Importantly, as both terms in Eq. \ref{eq:fast_spectroscopy_signal} contain $\lambda(f)$, the pure spectral response can be recovered by varying $T$ in post-processing (\ref{sec:bayesian_signal_analysis}). In particular, for the Lorentzian spectral response:
\begin{equation}\label{eq:lorentzian_spectral_response}
    \lambda_\mathrm{L}(f) = \Co \frac{(\frac{\Gamma}{2})^2}{f^2 + (\frac{\Gamma}{2})^2}\,,
\end{equation}  
with FWHM $\Gamma$ and on-resonance brightness $\Co$, the signal converges to: $\Cmean(f) \rightarrow\lambda_\mathrm{L}(f-f_1)$ at high threshold values ($T\gg \max\left[\lambda(f)\right]$), simplifying the analysis (\ref{sec:bayesian_signal_analysis}).

We demonstrate the check-probe spectroscopy method on the same $\vtwo$ center as used in Figs. \ref{fig:intro}f, \ref{fig:resonance_check} and \ref{fig:spectral_diffusion_dynamics} (Methods), and observe narrow Lorentzian resonances around the $f_1$ laser frequency ($\SI{39(3)}{\mega \hertz}$ at $T=7$, see Fig. \ref{fig:lifetime_limited}b). Correcting for residual broadening by fitting Eq. \ref{eq:fast_spectroscopy_signal} to the data for $1 \leq T \leq 13$ (using Eqs. \ref{eq:spectral_probability_density} and \ref{eq:lorentzian_spectral_response}), we extract: $\Co = 6.5(1)$ counts ($\SI{3.22(7)}{\kilo \hertz}$ count rate) and $\Gamma = \SI{33(1)}{\mega \hertz}$. This spectrum (as well as the those measured in Fig. \ref{fig:LZS_interference}) corresponds to an average over the $A_1$ and $A_2$ transitions, resulting in: $\Gamma \approx (\Gamma_\mathrm{A_1} + \Gamma_\mathrm{A_2})/2 $ ($< 5 \%$ deviation, assuming equal initialisation probability, see \ref{sec:mean_linewidth}). Note that this ambiguity between the transitions can be fully resolved by executing the `check' block with two NIR lasers, as in Fig. \ref{fig:resonance_check}. The discrepancy between the extracted mean linewidth and the mean lifetime limit ($\sim \SI{20}{\mega \hertz}$) is well-explained by power-broadening, with optical Rabi frequencies estimated to be $\sim \SI{26}{\mega \hertz}$ (next section, see Fig. \ref{fig:LZS_interference}j).

Next, to verify our previous inference that spectral diffusion is virtually absent without laser illumination (Fig. \ref{fig:spectral_diffusion_dynamics}c), we insert a \SI{40}{\milli \second} waiting time between the $f_1$ and $f_2$ pulses (Fig. \ref{fig:lifetime_limited}b, top), which does not increases the linewidth within the fit error ($T = 7$). Importantly, this allows for the preparation of the transition at a specific frequency, `storing' it in the dark, so that the $\vtwo$ center can be used to produce coherent photons at a later time. Furthermore, the broad nature of the diffusion-averaged linewidth depicted in Fig. \ref{fig:intro}f, enables probabilistic tuning of the emission frequency over more than a gigahertz \cite{brevoord_heralded_2024}. We demonstrate this by varying the $f_1$ frequency, initialising the emitter at different spectral locations, and probing the transition with the NIR laser at frequency $f_2$ (Fig. \ref{fig:lifetime_limited}c, d and e). Such tuning of the $\vtwo$ emission frequency without the need for externally applied electric fields \cite{ruhl_stark_2020, lukin_spectrally_2020} might open up new opportunities for optically interfacing multiple centers.

\subsection*{Check-probe spectroscopy: Landau-Zener-St{\"u}ckelberg interference}
\begin{figure*}[ht]
  \includegraphics[width=1 \textwidth]{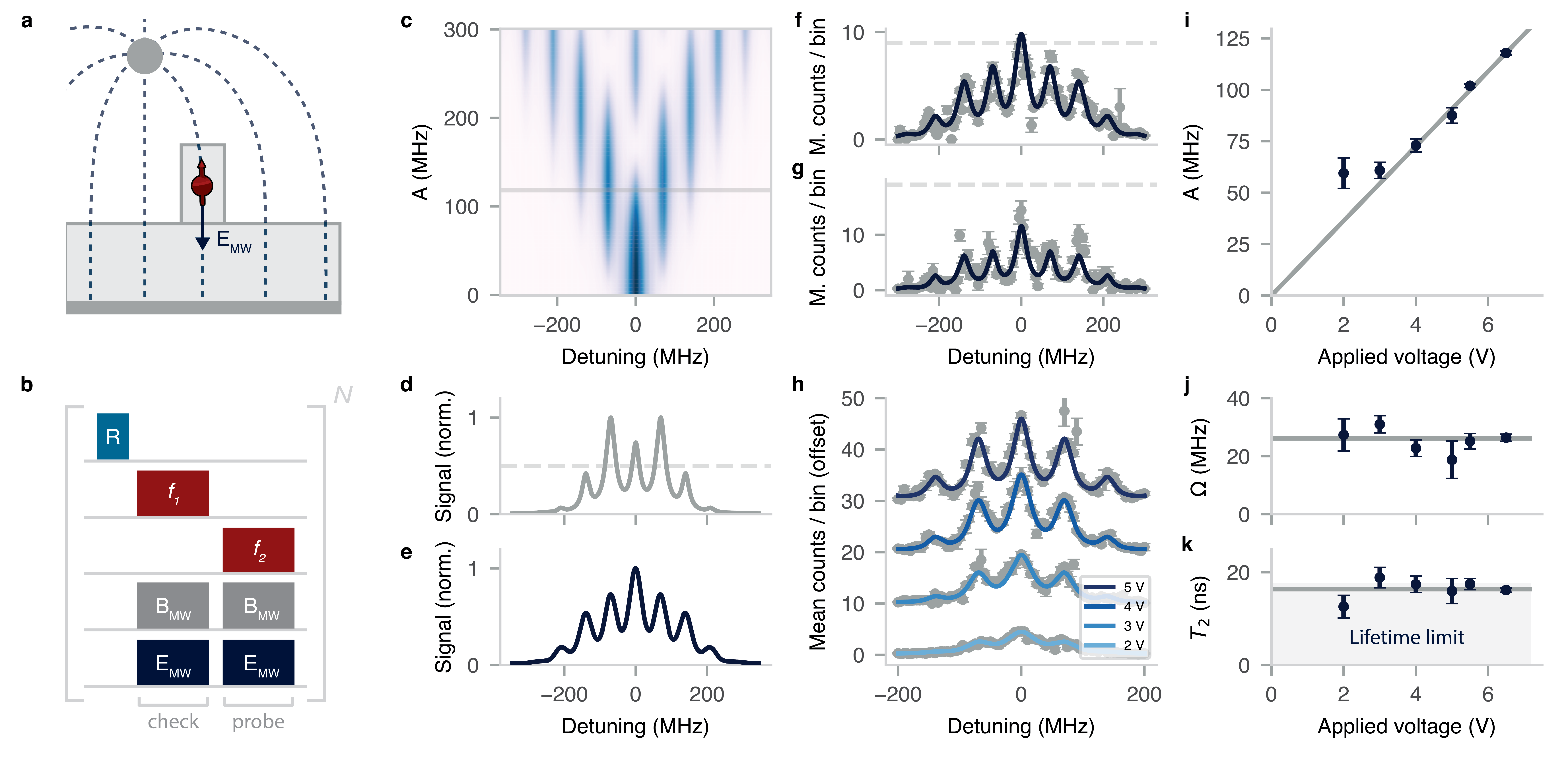}
  \caption{\label{fig:LZS_interference} \textbf{Landau-Zener-St{\"u}ckelberg (LZS) interference.} \textbf{a)} Schematic showing the electric-field generated by the MW drive, connecting the bond wire and ground plane (dark grey, not drawn to scale), which can generate significant stark shifts. \textbf{b)} Sequence, as in Fig. \ref{fig:lifetime_limited}a, but now explicitly including the electric field (E\textsubscript{MW}). \textbf{c)} Characteristic LZS interference pattern (theory) as function of the laser detuning, and the electric field strength $A$. At higher electric fields, side bands emerge at multiples of the driving frequency $\omega = \SI{70}{\mega \hertz}$. \textbf{d)} Line cut through (c) for $A = \SI{118}{\mega \hertz}$. The dashed line denotes an example threshold $T$. \textbf{e)} The experimental spectrum expected from the situation in (d). The signal differs from the original spectrum as it is weighted over the probability to pass the threshold for different detuning. \textbf{f)} Mean detected counts as a function of the two-laser detuning when the MW amplitude is set to \SI{6.5}{\volt}, approximately equal to the value in (d). The threshold ($T=10$) is set to about half the maximum amplitude, as in (d). The fit function (solid line) is obtained by fitting the dataset for a range of threshold values (see Supplementary Fig. \ref{fig:SI_LZS_2D_fits}). \textbf{g)} Same dataset as in (f), but with $T=20$. The signal distortion due to the threshold is well-captured by the fit. \textbf{h)} Experimental data (grey) and fit (solid lines) as in (f), varying the applied voltage ($T = \{ 6,10,13,15\}$). Data are offset by 10 counts for clarity. \textbf{i)} Extracted electric field strength $A$ as a function of the applied voltage. Solid grey line is a linear guide to the eye. \textbf{j)} Extracted optical Rabi frequency $\Omega$. Solid grey line denotes the inverse-variance weighted mean. \textbf{k)} Extracted optical coherence time $T_2$. Shaded region denotes the mean lifetime limit: $T_2 = 2 T_1 \approx \SI{17}{\nano \second}$ Solid grey line denotes the inverse-variance weighted mean of the data points. 
  }
\end{figure*}
To further benchmark the check-probe spectroscopy method, we use it to resolve Landau-Zener-St{\"u}ckelberg (LZS) interference fringes in the optical spectrum \cite{shevchenko_landau_2010}. Such fringes demonstrate coherent control of the orbital states of the V2 defect using MW frequency electric fields, and enable the independent determination of the optical coherence and Rabi frequency \cite{shevchenko_landau_2010}, allowing for the separation of their contributions to the linewidths measured in Fig. \ref{fig:lifetime_limited}.

LZS interference fringes can arise when a strong AC electric field shifts the optical transition across the laser frequency multiple times within the coherence time of the emitter \cite{miao_electrically_2019, lukin_spectrally_2020}. Each time a crossing occurs, the emitter is excited with a small probability amplitude and associated `St{\"u}ckelberg' phase. These amplitudes can interfere constructively or destructively, creating fringes in the spectrum (see Ref. \cite{shevchenko_landau_2010} for an extensive review on the phenomenon).

In our setup, MW radiation is applied by running an AC current through an aluminum alloy wire spanned across the sample (Fig. \ref{fig:LZS_interference}a). The original purpose of the wire is to enable mixing of the ground-state spin in the `check' and `probe' blocks used in Fig. \ref{fig:lifetime_limited}. However, this geometry also creates significant \emph{electric} fields at microwave frequencies (Fig. \ref{fig:LZS_interference}a). Taking the defect ground and excited states as basis states ($\ket{g} = \ket{0}$ and $\ket{e}=\ket{1}$), the (optical) evolution of the system is described by the Hamiltonian (in the rotating frame of the emitter) \cite{miao_electrically_2019}:
\begin{equation} \label{eq:lzs_hamiltonian}
    H = \frac{\Omega}{2} \,\sigma_x + \frac{\delta + A \, \cos{\left(\omega \, t\right)}}{2} \, \sigma_z \, ,
\end{equation}
where $\Omega$ is the optical Rabi frequency, $\delta$ is the detuning between the optical transition and the laser frequency, $A$ is the stark-shift amplitude (which scales with the electric field amplitude), $\omega$ is the MW driving frequency and $\sigma_x$, $\sigma_z$ are the Pauli spin matrices. For the experimental parameters used here, the system is considered to be in the `fast-passage' regime (defined as $A\,\omega \gg \Omega^2$ \cite{shevchenko_landau_2010}), meaning the excitation probability amplitude during a single crossing is small (Methods). In this regime, the spectral response of the emitter can be described by \cite{shevchenko_landau_2010}:

\begin{equation} \label{eq:lzs_solution}
    \lambda_{\mathrm{LZS}}(f) = \Co \sum_k \frac{\Omega^2_k}{\frac{1}{T_1 T_2} + \frac{T_2}{T_1}\left(k \, \omega - f\right)^2 + \Omega^2_k} \,,
\end{equation}
where $\Co$ is the on-resonance emitter brightness for $A=0$, $\Omega_k = \Omega \, J_k \left(\tfrac{A}{\omega}\right)$, with $J_k$ the Bessel function, and $T_1$, $T_2$ are the emitter's optical relaxation and (pure-dephasing) coherence times respectively. Figure \ref{fig:LZS_interference}c shows the optical spectrum as a function of $A$, obtained by evaluating Eq. \ref{eq:lzs_solution} for our sample parameters. At higher electric field amplitudes (i.e. higher $A$), multiple characteristic interference fringes appear, spaced by the driving frequency $\omega$ (here \SI{70}{\mega \hertz}), creating a complex optical spectrum. 

Measuring such complex spectra with the check-probe optical spectroscopy method requires taking into account signal distortions arising from the form of the spectral probability density after the check block (Eq. \ref{eq:spectral_probability_density}). To see why this is the case, we consider an exemplary theoretical spectrum plotted in Fig. \ref{fig:LZS_interference}d (for $A = \SI{118}{\mega \hertz}$), where the threshold $T$ is set to about half the maximum signal amplitude (dashed line in Fig. \ref{fig:LZS_interference}d). Such a threshold is passed (with high probability) not only when the central peak is on resonance with $f_1$, but also when one of the nearest fringes is on resonance with the laser. Computing the resulting weighted signal (inserting Eq. \ref{eq:lzs_solution} in Eq. \ref{eq:fast_spectroscopy_signal}) yields the distorted, experimentally expected spectrum shown in Fig. \ref{fig:LZS_interference}e. 


To experimentally measure the LZS interference signal, we execute the sequence in Fig. \ref{fig:LZS_interference}b, now explicitly including the electric field components generated by the MW drive. These components were also implicitly present in previously discussed experiments (Fig. \ref{fig:lifetime_limited}), but their effects could largely be neglected under the conditions: $\omega > A$ and $\omega > \Gamma \approx \sqrt{(\pi T_2)^{-2} + \Omega^2}$ \cite{shevchenko_landau_2010}. We set the MW driving frequency $\omega$ to  the ground-state zero field splitting (\SI{70}{\mega 
\hertz}), so that the magnetic field components efficiently mix the spin states \cite{banks_resonant_2019, babin_fabrication_2022}, and set the (peak-to-peak) MW amplitude between the wire and the ground plane to \SI{6.5}{\volt}. Figures \ref{fig:LZS_interference}f and g show the measured spectrum for a threshold of $T=10$ and $T=20$ respectively. The former corresponds roughly to the example threshold in Fig. \ref{fig:LZS_interference}d, and the observed signal matches well with the expected spectrum in Fig. \ref{fig:LZS_interference}e. Setting $T=20$ alters the measured spectrum, highlighting the interplay between the threshold and corresponding distortion. The solid lines are generated by a single fit of the complete dataset using Eq. \ref{eq:fast_spectroscopy_signal} for $1 < T < 21$ (post-processed, see Supplementary Fig. \ref{fig:SI_LZS_2D_fits}). 

We repeat this procedure while varying the MW amplitude (Fig \ref{fig:LZS_interference}h) and extract $A$, $\Omega$ and the estimated pure-dephasing $T_2$ coherence time (Figs \ref{fig:LZS_interference}i,j and k), keeping $\omega$ and the optical relaxation time $T_1 =\SI{8.7}{\nano \second}$ fixed (again using the mean of the $A_1$ and $A_2$ transitions \cite{liu_silicon_2024}). For amplitude values above \SI{2}{\volt} we observe a linear relation between the MW amplitude and $A$, as expected. For lower values, a significant deviation is observed, possibly because the system is no longer well-described by the fast-passage limit (i.e. $A\,\omega \sim \Omega^2$). Indeed, the optical Rabi frequency is estimated to be \SI{26(1)}{\mega \hertz} (weighted mean of Fig. \ref{fig:LZS_interference}j), on the order of $\sqrt{A\,\omega}$. Finally, we find a mean $T_2 =\SI{16.4(4)}{\nano \second}$, approximately equal to the mean lifetime limit ($2\,T_1 \approx \SI{17}{\nano \second}$ \cite{liu_silicon_2024}). 

\section*{Discussion}
In this work, we introduced a high-bandwidth check-probe scheme that allows for quantitative characterisation of spectral diffusion and ionisation processes under the influence of external perturbations. Our methods enable measurements of the homogeneous transition linewidth of single quantum emitters, under minimal system disturbance. 

We applied these methods to study the optical coherence of the $\vtwo$ center in commercially available bulk-grown silicon carbide. Despite high levels of spectral diffusion under laser illumination, we reveal near-lifetime-limited linewidths with slow dynamics, enabling the preparation of a frequency-tunable coherent optical transition \cite{brevoord_heralded_2024,ruhl_stark_2020,lukin_spectrally_2020}. Although higher purity materials are likely desired, such coherent optical transitions in bulk-grown SiC might enable nanophotonic device development, testing and characterisation (e.g. cavity coupling, Purcell enhancement) using widely available materials \cite{lukin_4hsiliconcarbideoninsulator_2020}. Future avenues for research in bulk-grown material include investigating the spin coherence properties and the spectral stability under higher-power laser pulses used for long-distance entanglement generation \cite{pompili_realization_2021,hermans_qubit_2022}. 

Finally, the presented methods are applicable to other platforms where spectral diffusion forms a natural challenge, such as rare-earth doped crystals \cite{thiel_rareearthdoped_2011}, localised excitons \cite{iff_substrate_2017} or semiconductor quantum dots \cite{neuhauser_correlation_2000}, and might enable new insights in the charge environment dynamics of such systems. 
\bibliography{references_SiC}
\section*{Methods}

\subsection*{Sample parameters}
The sample was diced directly from a 4-inch High-Purity Semi-Insulating (HPSI) wafer obtained from the company Wolfspeed, model type W4TRF0R-0200. We note that the HPSI terminology originates from the silicon carbide electronics industry. In the quantum technology context considered here, this material has a significant amount of residual impurities (order $\sim10^{15}\,\SI{}{\per\centi\meter\cubed}$ according to Son et al.\cite{son_charge_2021}) and is hence considered low purity with respect to a concentration of $\sim10^{13}\,\SI{}{\per\centi\meter\cubed}$ typical for epitaxially grown layers on the c-axis of silicon carbide \cite{nagy_highfidelity_2019,heiler_spectral_2024}. On a different sample, diced from a wafer with the same model type, a Secondary-Ion Mass Spectroscopy (SIMS) measurement determined the concentration of nitrogen donors as [N] = $\SI{1.1e15}{\per\centi\meter\cubed}$. In addition to intrinsic silicon vacancies, we generate additional silicon vacancies through a 2 MeV electron irradiation with a fluence of $\SI{5e13}{\per\centi\meter\squared}$. The sample was annealed  at \SI{600}{\celsius} for 30 min in an Argon atmosphere. To enhance the optical collection efficiency and mitigate the unfavourable $\vtwo$ dipole orientation for confocal access along the SiC growth axis (c-axis), we fabricate nanopillars. We deposit $\SI{25}{\nano\meter}$ of $\mathrm{Al_2O_3}$ and  $\SI{75}{\nano\meter}$ of nickel on lithographically defined disks. A subsequent $\mathrm{SF_6/O_2}$ ICP-RIE etches the pillars, see figure \ref{fig:intro}c. The nanopillars have a diameter of $\SI{450}{\nano\meter}$ at the top and $\SI{650}{\nano\meter}$ at the bottom and are $\SI{1.2}{\micro\meter}$ high. Considering the modest efficiency of our detector ($\approx 25 \%$ at \SI{950}{\nano \meter}), the $\vtwo$ centers studied here appear to be significantly brighter than those in epitaxially grown layers \cite{babin_fabrication_2022, lukin_twoemitter_2023, banks_resonant_2019, korber_fluorescence_2024}, which is consistent with previous studies in commercially available material \cite{radulaski_scalable_2017, widmann_coherent_2015}. 

\subsection*{Experimental setup}
All experiments are performed using a home-built confocal microscopy setup at 4K (Montana Instruments S100). The NIR lasers (Toptica DL Pro and the Spectra-Physics Velocity TLB-6718-P) are frequency-locked to a wavemeter (HF-Angstrom WS/U-10U) and their power is modulated by acousto-optic-modulators (G\&H SF05958). A wavelength division multiplexer (OZ Optics) combines the 785 nm repump (Cobolt 06-MLD785) and NIR laser light, after which it is focused onto the sample by a movable, room temperature objective (Olympus MPLFLN 100x), which is kept at vacuum and is thermally isolated by a heat shield. A 90:10 beam splitter that directs the laser light into the objective, allows $\vtwo$ center phonon-sideband emission to pass through, to be detected on an avalanche photodiode (COUNT-50N, filtered by a Semrock FF01-937/LP-25 long pass filter at a slight angle). Alternatively, emission can be directed to a spectrometer (Princeton Instruments IsoPlane 81), filtered by a 830 nm long pass filter (Semrock BLP01-830R-25). Microwave pulses are generated by an arbitrary-waveform generator (Zurich Instruments HDAWG8), amplified (Mini-circuits LZY-22+), and delivered with a bond-wire drawn across the sample. The  coarse time scheduling  ($\SI{1}{\micro \second}$ resolution) of the experiments is managed by a microcontroller (ADwin Pro II). For a schematic of the setup see Supplementary Fig. \ref{fig:SI_Optical_Setup}.

\subsection*{Magnetic field}
For the check-probe optical spectroscopy measurements in Fig. \ref{fig:lifetime_limited} (and Supplementary Fig. \ref{fig:SI_Fast_PLE_plot}), we apply an external magnetic field of $\approx\SI{1}{\milli \tesla}$ along the defect symmetry axis (c-axis). All other experiments are performed at approximately zero field. We apply the field by placing a permanent neodymium magnet outside the cryostat. We align it by performing the sequence depicted in Fig. \ref{fig:intro}e, with $f_1$ set at the center frequency of the broad resonance (Fig. \ref{fig:intro}f), and monitoring the average photoluminescence ($f_1$ pulse is set to \SI{2}{\milli \second}). A (slightly) misaligned field causes spin-mixing between the $m_s = \pm \tfrac{3}{2}$ and $m_s=\pm \tfrac{1}{2}$ subspace, which increases the detected signal. Minimising for the photoluminescence thus optimises the field alignment along the symmetry axis. 

\subsection*{LZS fast-passage regime}
The fast-passage regime is defined by: $A \omega \gg \Omega^2$ \cite{shevchenko_landau_2010}, with $\omega = \SI{70}{\mega \hertz}$. From the measurements in Fig. \ref{fig:lifetime_limited}, we can estimate: $\Omega < \SI{40}{MHz}$. Furthermore, we can get a rough estimate for $A$ by approximating the electric field at the defect to be: $E \approx \tfrac{U}{d}(\epsilon + 2)/3 $, with $U$ the applied voltage, $d \approx \SI{500}{\micro \meter}$ the distance between the wire and the ground plane and $\epsilon \approx 10$ the dielectric constant of silicon carbide (using the local field approximation \cite{lukin_spectrally_2020}). There is some debate on the value of the Stark-shift coefficient \cite{ruhl_stark_2020, lukin_spectrally_2020}. Here, we take the value from Ref. \cite{lukin_spectrally_2020}: \SI{3.65}{\giga \hertz \meter \per \mega \volt} and estimate $A\approx \SI{29}{\mega \hertz}$ for an applied voltage of \SI{1}{\volt}. Hence, $A \omega > \Omega^2$ for $U > \SI{1}{\volt}$, and the fast-passage requirement is satisfied for most measurements in Fig. \ref{fig:LZS_interference}. The excellent agreement between the data and our model (especially for higher values of $U$) and the corresponding extracted values for $\Omega$ and $A$, further justify using the fast-passage solution of the LZS Hamiltonian. 

\subsection*{Error analysis}
For all quoted experimental values, the value between brackets indicates one standard deviation or the standard error obtained from the fit (unless stated otherwise). The error bars on the mean counts are based on Poissonian shot noise. The uncertainty on fit parameters is rescaled to match the sample variance of the residuals after the fit. 
\section*{Acknowledgements}
We thank Florian Kaiser and Michel Orrit for useful discussions relating to the spectral diffusion measurements. We thank Tien Son Nguyen, Michael Flatté and Denis Candido for input on the charge stability and charge dynamics. We thank Benjamin Pingault for his input on the design of the experimental setup and Daniel Bedialauneta Rodriguez for his input during project discussions. We thank Yanik Herrmann for feedback on the manuscript. We thank Nico Albert and Tim Hiep for designing and machining parts for the setup, Henri Ervasti, Pieter Botma and Ravi Budhrani for software support and development, Régis Méjard and Hitham Mahmoud Amin for optical and safety support, Siebe Visser, Vinod Narain and Erik van der Wiel for general technical support and Jason Mensigh, Olaf Benningshof and Jared Croese for cryogenic and vacuum engineering support.

This research was supported by the Quantum Internet Alliance through the European Union’s Horizon Europe program grant agreement No. 101080128. This project has received funding from the European Research Council (ERC) under the European Union’s Horizon 2020 research and innovation programme (grant agreement No. 852410). This work was supported by the Dutch National Growth Fund (NGF), as part of the Quantum Delta NL programme. This work is part of the research programme NWA-ORC with project number NWA.1160.18.208, which is (partly) financed by the Dutch Research Council (NWO). This project has received funding from the European Union’s Horizon Europe research and innovation program under grant agreement No
 101135699.

\section*{Author contributions}
GLvdS, LJF, SJHL and THT devised the experiments. GLvdS, LJF and AD performed the experiments and collected the data. GLvdS, LJF and SJHL prepared the experimental apparatus. GLvdS, LJF, SJHL and THT analysed the data. GMT and TWdJ designed and fabricated the sample. GLvdS, LJF, SJHL and THT wrote the manuscript with input from all authors. THT supervised the project.

\section*{Data availability}
All data underlying the study will be made available on the open 4TU data server.

\section*{Code availability}
Code used to operate the experiments is available on request.

\section*{Competing interests}
The authors declare no competing interests.

\clearpage
\newpage






\clearpage
\widetext
\begin{center}
\textbf{\large Supplementary Information for \\``Check-probe spectroscopy of lifetime-limited emitters in bulk-grown silicon carbide''}
\end{center} 

\renewcommand{\figurename}{\textbf{Supplementary Figure}}
\renewcommand{\tablename}{\textbf{Supplementary Table}}
\renewcommand{\thesubsection}{Supplementary Note \arabic{subsection}}
\setcounter{subsection}{0}

\setcounter{equation}{0}
\setcounter{figure}{0}
\setcounter{table}{0}
\setcounter{page}{1}
\makeatletter
\renewcommand{\theequation}{S\arabic{equation}}


\subsection{Spectral dynamics model derivation}
\label{sec:model_derivation}

The Lorentzian spectral propagator is given by \cite{zumofen_spectral_1994}:
\begin{equation} \label{eq:spectral_propagator}
    P_f(t) = \pi^{-1} \frac{\frac{1}{2}\gamma(t)}{ f^2 + \left(\frac{1}{2}\gamma(t)\right)^2} \, ,
\end{equation}
where $f$ is the detuning from resonance.  The FWMH $\gamma(t)$ is linear in time \cite{zumofen_spectral_1994}:
\begin{equation} \label{eq:linear_fwhm}
    \gamma(t) = \gdif \, t.
\end{equation}

The mean detected counts a function of detuning $f$ between the laser and the emitter's resonance frequency is given by a Lorentzian:
\begin{equation}
    \lambda_\mathrm{L}(f) = \Co   \, \frac{\left( \frac{\Gamma}{2}\right)^2}{ f^2 + \left(\frac{\Gamma}{2}\right)^2} \, ,
\end{equation}
with $\Co$ the observed counts on resonance and $\Gamma$ the emitters homogeneous line width (FWHM). This Lorentzian is normalised such that $\lambda(0) = \Co$. We can add the effect of ionisation by describing the ionisation probability as:
\begin{equation}
    P_{\mathrm{i}}(t) = 1 - e^{-\gion t} \, ,
\end{equation}
Here, $\gion$ is assumed to be independent of emitter detuning. However, if ionisation is caused by a two-photon process, we would expect the ionisation rate to scale with the square of the excitation rate:
\begin{equation} \label{eq:ionisation_region}
    \gion(f) = \giono \, \left(\frac{(\frac{1}{2}\Gamma)^2}{f^2 + (\frac{1}{2}\Gamma)^2} \right)^2 ,
\end{equation}
where $\giono$ is the ionisation rate on resonance.  Then,
\begin{equation}
    \gion = \int_{-\infty}^\infty P_f \, \gion(f) df = \frac{\gamma_i^0}{2} \frac{\Gamma \gamma(t) + 2 \Gamma^2}{(\gamma(t) + \Gamma)^2} 
    \label{SI:eq_gamma_resonant}
\end{equation}

The mean counts can then be calculated as:
\begin{equation}
    \lambda(f,t) = \left(1-P_\mathrm{i}(t) \right) \, \lambda_\mathrm{L}(f) \, ,
\end{equation}
Next, we integrate over the spectral density:
\begin{equation}
    \Cmean(t) = \int_{-\infty}^\infty P_f(t) \, \lambda(f,t) \, df,
\end{equation}
which can be solved to yield:
\begin{equation}
    \Cmean(t) = \Co \frac{1}{1 + \gdif \, t /\Gamma} e^{-\gion t}.
\end{equation}
\noindent where we can either decide to model $\gamma_i$ as a constant, or by equation \eqref{SI:eq_gamma_resonant}. Additionally, if we neglect the recapture or de-ionisation rate, then for values of $t<0$, effectively $\gamma_i=0$.

\clearpage
\subsection{Spectral dynamics model performance and dependence on set counts threshold}
\label{sec:threshold_dependence_rates}
The model presented in the main text describes spectral diffusion caused by a macroscopic bath of fluctuating dipoles (described by the spectral propagator in Eq. \ref{eq:spectral_propagator}). We can compare the performance of this model compared to other versions of the model, which restric certain parameters, for example one that assumes that no spectral diffusion (or ionisation) occurs. Supplementary Table \ref{tab:diffusion_models} presents the description of the models we fitted to the data.

Sweeping the counts threshold value $T$, we observe that all models saturate to a range of values for higher $T$, consistent with the intuition that stringent thresholds initialise the emitter more precisely on resonance with the lasers. At very high tresholds ($T\gtrapprox 35$), only very little data is available, and rates obtained with all models start to fluctuate heavily (due to the noisy data). Hence, we emperically determine a convergence region: $20 \leq T \leq 27$ (grey solid lines in Supplementary Fig. \ref{fig:SI_sweep_threshold} and $12 \leq T \leq 21$ in Supplementary Fig. \ref{fig:SI_sweep_threshold_offres}), over which we average to obtain the rates presented in the main text.

Comparing various goodness of fit metrics (reduced $\chi$-squared, AIC, BIC), we learn that the `No recap' model describes the data best (Supplementary Fig. \ref{fig:SI_sweep_threshold}j-l), indicating that recapture is virtually not present under NIR-laser illumination.

\begin{figure*}[ht]
  \includegraphics[width=1 \columnwidth]{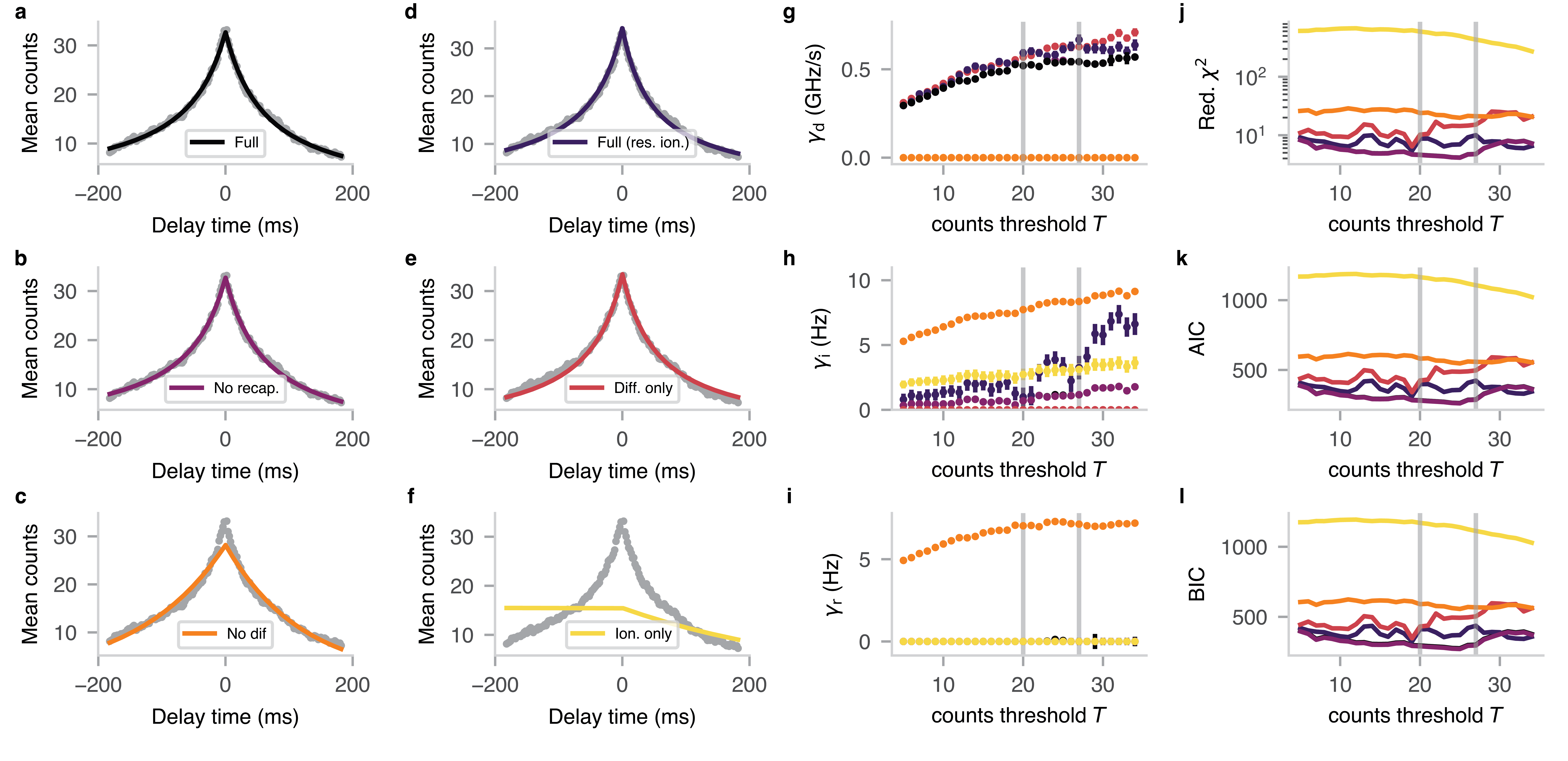}
  \caption{\label{fig:SI_sweep_threshold} \textbf{Comparison of models for perturbation by the NIR lasers.} \textbf{a-f)} Data from Fig. \ref{fig:spectral_diffusion_dynamics}d, overlayed with the various models, which are described in Supplementary Table \ref{tab:diffusion_models}. \textbf{g-i)} Estimated rates $\gdif, \giono, \grecap$ as a function of counts threshold $T$ for the different models. We observe a saturating behaviour for high $T$, as the defect is initialised in a more specific frequency range. For very high treshold values, the lack of data creates large fluctuations in the obtained rates. To determine the rate values in Supplementary Table \ref{tab:diffusion_models}, we take an average over the values indicated by the grey solid lines and take the standard deviation as uncertainty. \textbf{j-l)} Various goodness of fit models (reduced $\chi$-squared, AIC, BIC) indicate that the model presented in the main text (Std.) captures the data best. Here, we also observe a saturation behaviour, where the sudden drop for very high $T$ is attributed to overfitting due to lack of data. Colors in (g-l) match the models described in (a-f). The `Full' data overlaps with the `Full (res.ion.)' curve. Errorbars are based on photon shot noise (grey data), or represent fit uncertainties (panels g-i).}
\end{figure*}

\begin{figure*}[ht]
  \includegraphics[width=1 \columnwidth]{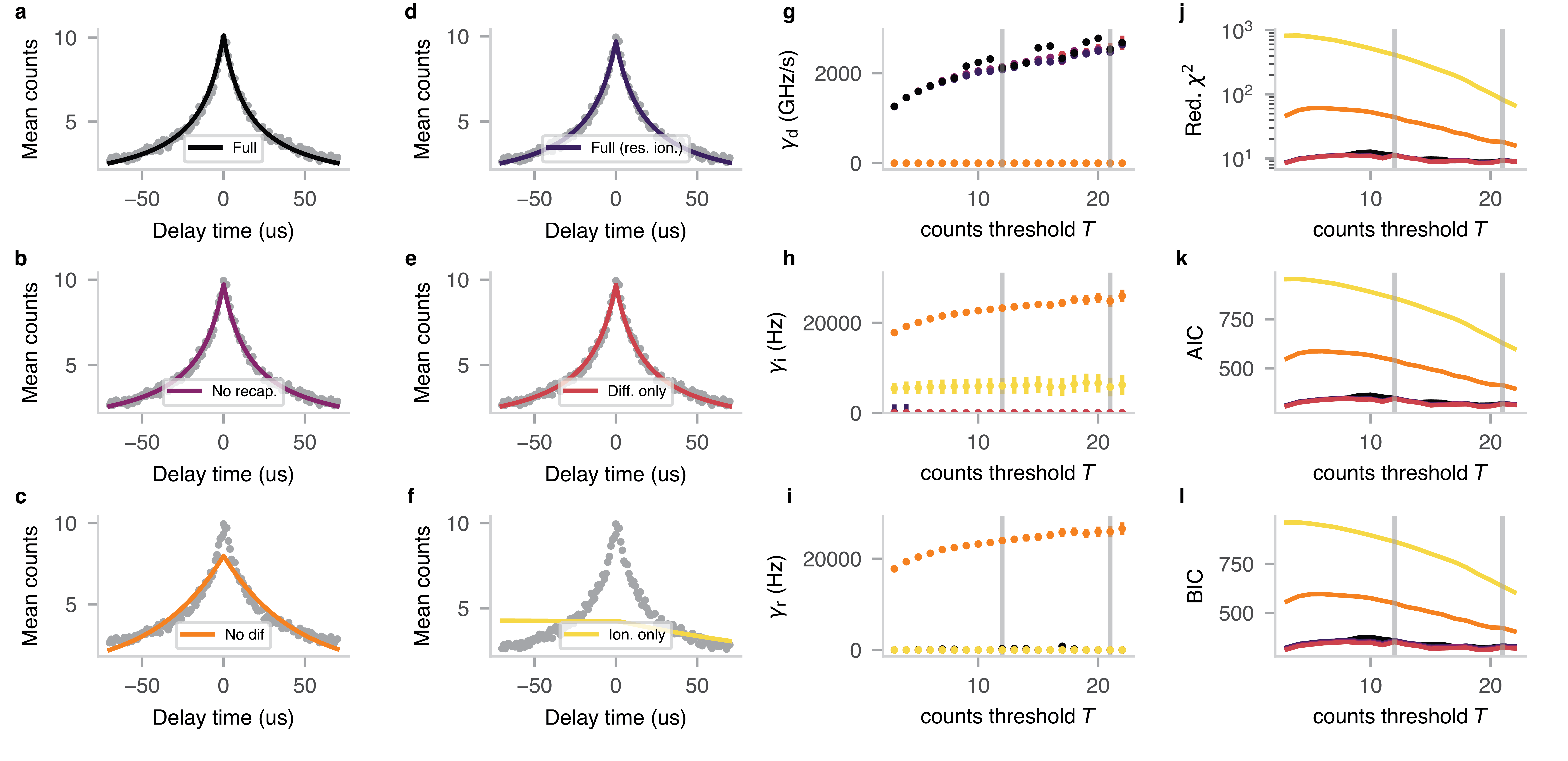}
  \caption{\label{fig:SI_sweep_threshold_offres} \textbf{Comparison of models for perturbation by the repump laser.} \textbf{a-i)} Same as Supplementary Fig.\ref{fig:SI_sweep_threshold}, but for the data presented in Fig. \ref{fig:spectral_diffusion_dynamics}e. }
\end{figure*}


\begin{table}[ht] 
    \setlength\tabcolsep{9pt}

    \def\arraystretch{2}

    \centering
    \caption{Models used for fitting in Supplementary Figs. \ref{fig:SI_sweep_threshold} and \ref{fig:SI_sweep_threshold_offres}}
    \begin{tabular}{l|l}

    \toprule
    Model name  & Description \\
    \midrule
    \midrule

    Full & 
    $\Cmean(t)/\Co = \begin{cases}
    \left( 1 + \gdif \, t\, /\,\Gamma \right)^{-1} \,e^{-\gion t} , & \text{if $t > 0$}.\\
    \left( 1 - \gdif \, t\, /\,\Gamma \right)^{-1} \,e^{\grecap t} , & \text{otherwise}.
    \end{cases}$\\
   
    \midrule
    Full (res. ion.) &
    $\Cmean(t)/\Co = \begin{cases}
    \left( 1 + \gdif \, t\, /\,\Gamma \right)^{-1} \,e^{-\frac{\giono}{2} \frac{\Gamma \gamma(t) + 2 \Gamma^2}{(\gamma(t) + \Gamma)^2} t} , & \text{if $t > 0$}.\\
    \left( 1 - \gdif \, t\, /\,\Gamma \right)^{-1} \,e^{\grecap t}, & \text{otherwise}.
    \end{cases}$\\

    \midrule
    No recap. &
    $\Cmean(t)/\Co = \begin{cases}
    \left( 1 + \gdif \, t\, /\,\Gamma \right)^{-1} \,e^{-\gion t} , & \text{if $t > 0$}.\\
    \left( 1 - \gdif \, t\, /\,\Gamma \right)^{-1} , & \text{otherwise}.
    \end{cases}$\\

    \midrule
    Diff. only &
    $\Cmean(t)/\Co = \begin{cases}
    \left( 1 + \gdif \, t\, /\,\Gamma \right)^{-1} , & \text{if $t > 0$}.\\
    \left( 1 - \gdif \, t\, /\,\Gamma \right)^{-1}, & \text{otherwise}.
    \end{cases}$\\
    
    \midrule
    No diff. &
    $\Cmean(t)/\Co = \begin{cases}
    e^{-\gion t} , & \text{if $t > 0$}.\\
    e^{\grecap t}, & \text{otherwise}.
    \end{cases}$\\
    
    \midrule
    Ion. only &
    $\Cmean(t)/\Co = \begin{cases}
    e^{-\gion t} , & \text{if $t > 0$}.\\
    1, & \text{otherwise}.
    \end{cases}$\\

    \end{tabular}

    \label{tab:diffusion_models}
\end{table}





   



\clearpage
\subsection{Diffusion power dependence}
\begin{figure}[ht]
  \includegraphics[width=0.5 \columnwidth]{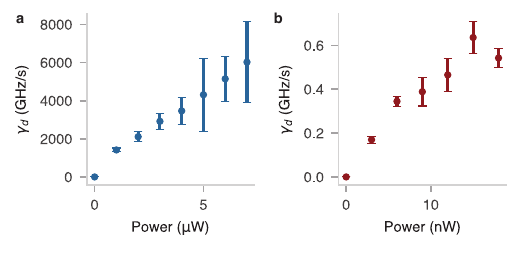}
  \caption{\label{fig:SI_laser_power_sweep} \textbf{Diffusion rates as function of laser power.} \textbf{a)} Diffusion rate $\gdif$ as a function of applied repump laser power (785 nm, as in Fig. \ref{fig:spectral_diffusion_dynamics}c). Data from 8-10 $\si{\micro \watt}$ are omitted as the CR-check was not passed, due to drifting of the objective (leading to a diminished collection efficiency). \textbf{b)} Diffusion rate $\gdif$ as a function of applied resonant laser power (916 nm, single laser). Data points for both panels are an average over fits with thresholds $T=\{10,14,18\}$, and error bars denote the standard deviation of the extracted rates. Laser powers are measured at the entrance of the objective. The rates presented in Fig. \ref{fig:spectral_diffusion_dynamics}d are an average of the two final (saturation) data points for both panels (with their standard deviation as error bar). }
\end{figure}

Here we plot the diffusion rates extracted by the method in figure \ref{fig:spectral_diffusion_dynamics} for different resonant and NIR laser powers. The data in figure \ref{fig:spectral_diffusion_dynamics} was averaged over the powers corresponding to the last two data points in figure \ref{fig:SI_laser_power_sweep}a and b for the repump and NIR laser respectively.

\begin{figure}[ht]
  \includegraphics[width=1 \columnwidth]{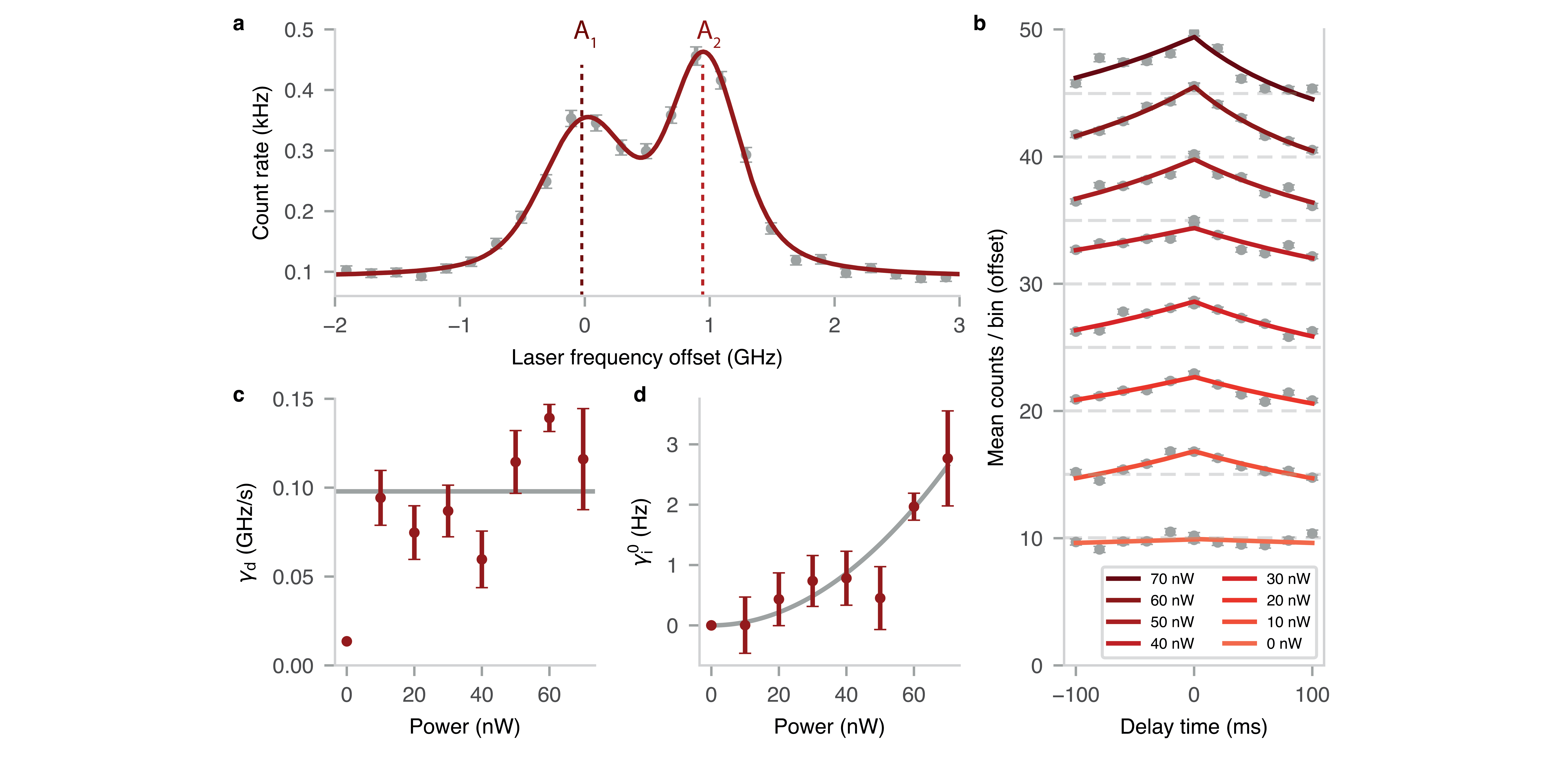}
  \caption{\label{fig:SI_laser_power_sweep_2} \textbf{Diffusion rates of a $\vtwo$ center in a different nanopillar} \textbf{a)} Diffusion-averaged PLE of the $\vtwo$ center in a different nanopillar (in a different sample) compared to Fig. \ref{fig:intro}f, yielding FWHMs of $\sim\SI{802}{\mega\hertz}$ and $\sim\SI{680}{\mega\hertz}$ for the $A_1$ and $A_2$ transitions, respectively.  \textbf{b)} Diffusion check-probe spectroscopy as in Fig. \ref{fig:spectral_diffusion_dynamics}d in the main text, for different NIR-laser powers (916 nm). We explicitly set $\grecap = 0$ (as suggested by Supplementary Fig. \ref{fig:SI_sweep_threshold}) to ensure convergence of the fit. The grey dotted lines indicate the threshold $T=10$ used for the `probe' block. Data are offset by 3 for visual clarity. \textbf{c)} Diffusion rates extracted from (b). with a mean diffusion rate of \SI{0.098}{\giga\hertz/\second} \textbf{d)} (On resonance) ionisation rates extracted from (b). Solid grey line is a guide to the eye: $y =  A\, x^2$, with $A= \SI{5.4e-4}{\hertz \per \nano \watt \squared}$ and $x$ the NIR laser power.}
\end{figure}

In another sample (same wafer and fabrication method, but with an irradiation dose of $\SI{2e12}{\per\centi\meter\squared}$), we investigated a $\vtwo$ center with significantly lower (saturation) spectral diffusion rates under NIR laser illumination ($\sim 6$ times smaller compared to the emitter from the main text, see Supplementary Fig. \ref{fig:SI_laser_power_sweep}). The extracted diffusion and ionisation rates for different NIR laser powers can be found in \ref{fig:SI_laser_power_sweep_2}b-d. We observe a saturation-type behaviour for spectral diffusion (up to \SI{70}{\nano\watt}) and an (approximately quadratic) increase in (on-resonance) ionisation rate for higher powers.   

\clearpage
\subsection{Bayesian analysis of the \emph{check-probe} spectroscopy signal} \label{sec:bayesian_signal_analysis}
The spectrum obtained from the \emph{check-probe} spectroscopy measurements in Figs. \ref{fig:lifetime_limited} and \ref{fig:LZS_interference} contains residual inhomogeneous broadening, due to non-perfect initialisation on-resonance with the $f_1$ laser frequency during the `check' block. To account for such residual broadening, we need to considere the spectral probability density of the emitter immediately after the `check' block, given that the minimum counts threshold $T$ was passed. Evidently, such a spectral probability density will strongly depend on $T$, with higher threshold values resulting in sharper distributions around $f_1$. To quantify this intuition, we apply Bayes' theorem to compute the spectral probability density after detecting $\mc \geq T$ photons:
\begin{equation} \label{eq:bayes_rule}
    P(f\,|\,\mc \geq T) = \frac{P(\mc \geq T\,| \, f) \, P(f)}{P(\mc)} = \frac{1}{N_T}P(\mc \geq T\,| \, f) \,,
\end{equation}
with $f$ the emitter frequency and $N_T$ some (numerically determined) normalisation constant. In the second step, we have assumed minimal knowledge on the spectral distribution before the `check' block, by taking the (improper) prior $P(f)$ to be (locally) flat. This assumption is justified if the diffusion-averaged inhomogeneous linewidth (Fig. \ref{fig:intro}f) is much larger than the homogenous linewidth (Fig. \ref{fig:lifetime_limited}b). Otherwise, a different prior function may be used (e.g. the Gaussian used for fitting the data in Fig. \ref{fig:intro}f).

Assuming Poissonian photon statistics, Eq. \ref{eq:bayes_rule} can be solved to yield:
\begin{equation} \label{eq:gamma_function}
    P(f\,|\,\mc \geq T) = \frac{1}{N_T} \sum_{T}^{\infty} \frac{\lambda(f-f_1)^{\mc} \, e^{-\lambda(f-f_1)}}{\mc !} = \frac{1}{N_T} \left(1 - \Gamma_\mathrm{i}\left[  T, \lambda(f-f_1) \right]\right) \,,
\end{equation}
with $\lambda(f)$ the expected mean number of counts during a single `check' block when the emitter is at frequency $f$ and the laser at frequency $f_1$, and $\Gamma_\mathrm{i}[a,z] = \frac{1}{\Gamma_\mathrm{c}[a]} \int_{z}^{\infty}t^{a-1}e^{-t}dt$, the incomplete Gamma function, with $\Gamma_\mathrm{c}[a]$ the Euler Gamma function. 

Here, $\lambda(f)$ encodes (our model of) the `pure' spectral response of the emitter, independent of residual inhomogeneous broadening, and solely governed by the underlying physical parameters whose value we aim to extract. For a simple Lorentzian lineshape, $\lambda(f)$ is given by:
\begin{equation} \label{eq:lambda_f_L}
    \lambda_{\mathrm{L}}(f) = \Co \frac{\left(\frac{\Gamma}{2}\right)^2}{f^2 + \left(\frac{\Gamma}{2}\right)^2}
\end{equation}
with $\Gamma$ the homogeneous linewidth (FWHM) and $\Co$ the mean number of detected counts on resonance. Conversely, the Landau-Zener-St\"uckelberg spectrum can be described by (see also Eq. \ref{eq:lzs_solution}):
\begin{equation} \label{eq:lambda_f_LZS}
    \lambda_{\mathrm{LZS}}(f) = \Co \sum_k \frac{\Omega^2_k}{\frac{1}{T_1 T_2} + \frac{T_2}{T_1}\left(k \, \omega - f\right)^2 + \Omega^2_k} \,,
\end{equation}
The spectral probability densities can then be computed by inserting $\lambda_\mathrm{L}$ or $\lambda_\mathrm{LZS}$ into Eq. \ref{eq:gamma_function}, the results of which are shown in Supplementary Fig. \ref{fig:SI_Bayesian_analysis}a and e, respectively. 

\begin{figure}[ht]
  \includegraphics[width=1 \columnwidth]{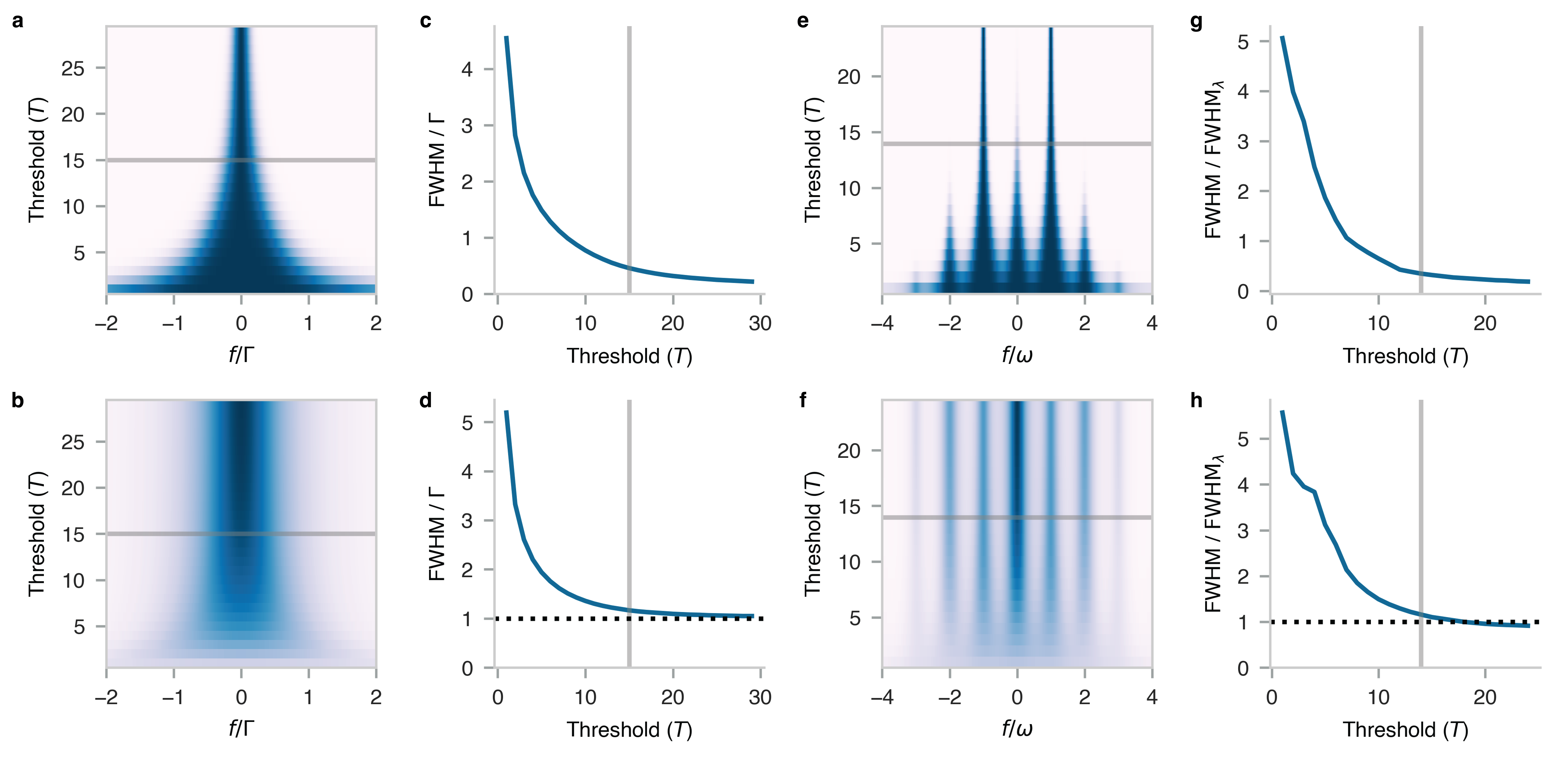}
  \caption{\label{fig:SI_Bayesian_analysis} \textbf{Bayesian spectrum analysis.} \textbf{a)} Spectral probability density $P(f\,|\,\mc \geq T)$ as a function of threshold for $\lambda(f) = \lambda_{\mathrm{L}}(f)$ (Lorentzian, Eq. \ref{eq:lambda_f_L}). Probabilities at each threshold value are normalised to their maximum value for visual clarity. Solid grey line denotes the maximum of $\lambda(f)$ for all panels. \textbf{b)} Check-probe spectroscopy signal, obtained by evaluating Eq. \ref{eq:convolution} for the spectral density calculated in (a). \textbf{c)} Full width at half maximum (FWHM) of the spectral density in (a). In the high-threshold regime ($T \gg \max{\left[\lambda(f) \right]}$, the linewidth becomes negligible compared to the linewidth of $\lambda(f)$ itself. \textbf{d)} Linewidth of the spectroscopy signal, which converges to that of the pure spectral response $\lambda(f)$ for high thresholds (black dotted line). \textbf{e)} Same as in (a), but for $\lambda(f) = \lambda_\mathrm{LZS}(f)$, which has two global maxima (Eq. \ref{eq:lambda_f_LZS}). LZS parameters are chosen to be equal to those extracted from the 6.5 V measurement in the main text (see Fig. \ref{fig:LZS_interference} and Supplementary Fig. \ref{fig:SI_LZS_2D_fits}). \textbf{f)} Same as in (b), but for the spectral density calculated in (e). \textbf{g)} FWHM of (e), computed by integrating the frequency ranges for which $P(f\,|\,\mc \geq T)>\frac{1}{2} \max{\left[ P(f\,|\,\mc \geq T)\right]}$ (and normalised to the range for which: $\lambda_\mathrm{LZS}(f) > \frac{1}{2}\max{\left[\lambda_\mathrm{LZS}(f)\right]}$). \textbf{h)} Same as in (d), but for the signal shown in (f). Note that for the LZS spectrum, which contains multiple global maxima, the FWHM does not neccesarily converge to the FWHM of $\lambda(f)$.
  }
\end{figure}

Finally, to obtain an expression for the spectroscopy signal (i.e. the mean number of detected photons $\Cmean$  during the probe block), we take the convolution of the spectral probability density (i.e. the residual inhomogeneous broadening) and the pure spectral response $\lambda(f)$ (assuming equal duration of the `check' and `probe' blocks):
\begin{equation} \label{eq:convolution}
    \Cmean(f) = P(f\,|\,\mc \geq T) \, * \, \lambda(f)  \, ,
\end{equation}
Importantly, both terms in the convolution involve $\lambda(f)$ (through Eq. \ref{eq:gamma_function}), and the first is strongly dependent on the chosen threshold value (see Supplementary Fig. \ref{fig:SI_Bayesian_analysis}a and e), allowing for the extraction of the pure spectral response by sweeping $T$ (in post-processing). Moreover, as expected, the degree of residual inhomogeneous broadening diminishes for higher threshold values (see Fig. \ref{fig:SI_Bayesian_analysis}c and d). 


\begin{figure}[ht]
  \includegraphics[width=1 \columnwidth]{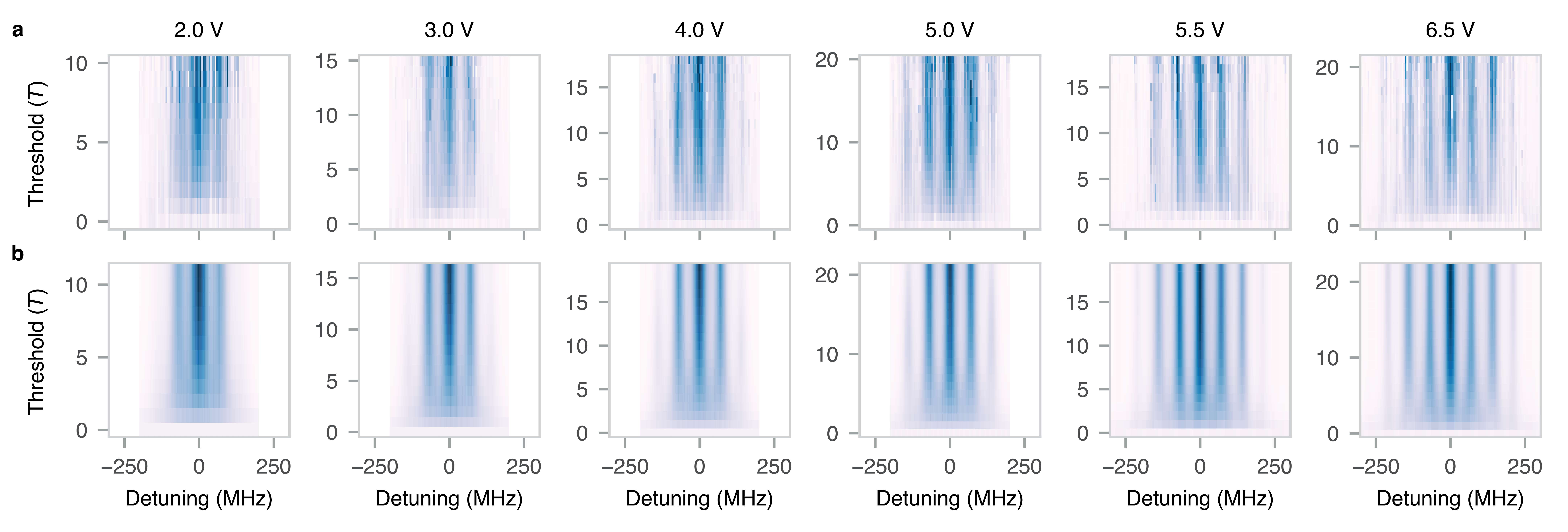}
  \caption{\label{fig:SI_LZS_2D_fits} \textbf{Bayesian analysis of the LZS interference signal.} \textbf{a)} Experimental data underlying the extracted parameters presented in Fig. \ref{fig:LZS_interference}i and j (applied voltage annotated), as a function of the two-laser detuning ($f_2 - f_1$) and the set threshold $T$. \textbf{b)} Fit of the data in (a) to Eq. \ref{eq:convolution}, using only three free fit parameters: $A, \Omega$ and $T_2$, together with a general scaling and offset. The fixed fit parameters $\omega$ and $T_1$ are set to \SI{70}{\mega \hertz} and \SI{7}{\nano \second}, respectively. The dependence on $T$ is fixed by the model (Eq. \ref{eq:gamma_function}).
  }
\end{figure}

\clearpage
\subsection{Mean linewidth approximation} \label{sec:mean_linewidth}
As noted in the main text, for the check-probe spectroscopy experiments that are executed with the MW mixing instead of the two NIR lasers, there is an approximately equal chance of initialising the system on the $A_1$ transition, or the $A_2$ transition. Hence, we expect the observed spectrum to be a (weigthed) average of both spectra, weighted by the respective initialisation probabilities. Assuming near-equal brightness of the transitions \cite{liu_silicon_2024}, we approximate this weighting to be roughly $50/50$.

\begin{figure}[h]
    \centering
    \includegraphics[width=0.66\linewidth]{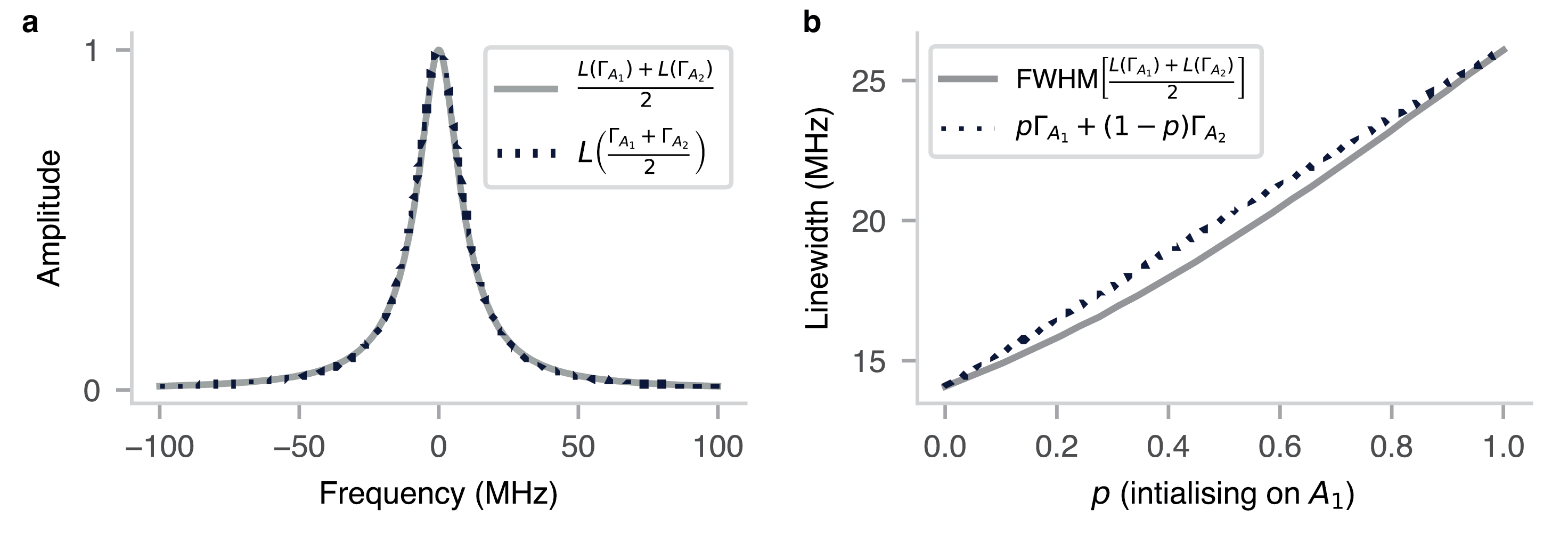}
    \caption{\textbf{Mean linewidth approximation.} \textbf{a)} Comparison between the equal mean of two Lorentzians with FWHM $\Gamma_\mathrm{A_1} = \SI{26}{\mega \hertz}$ and $\Gamma_\mathrm{A_2} = \SI{14}{\mega \hertz}$ (grey line) and a single Lorentzian function with a FWHM of $\Gamma = \tfrac{\Gamma_\mathrm{A_1}+\Gamma_\mathrm{A_2}}{2} = \SI{20}{\mega \hertz}$ (blue dotted line), showing qualitative agreement ($<1\%$ amplitude deviation). Legend: $L(\Gamma)$ denotes the Lorentzian distribution with FWHM $\Gamma$. \textbf{b)} Numerically determined linewidth (FWHM) of the weighted mean (with weights $p$ and $1-p$) of two Lorentzians with FWHM $\Gamma_\mathrm{A_1}$ and $\Gamma_\mathrm{A_2}$ as in (a) (grey line). The blue dotted approximates the actual linewidth by a (linear) weighted mean (Eq. \ref{eq:linewidth_weighted_mean}), showing deviations of less than $5\%$.   }
    \label{fig:mean_linewidth}
\end{figure}

To compare the measured linewidth (after accounting for residual inhomogeneous broadening, see \ref{sec:bayesian_signal_analysis}), with the expected lifetime limit, we approximate the weigthed mean of the two Lorentzian transitions, to be a Lorentzian with linewidth given by the weighted mean of the respective linewidths: 
\begin{equation}\label{eq:linewidth_weighted_mean}
    \Gamma = p\Gamma_\mathrm{A_1} + (1-p)\Gamma_\mathrm{A_2} \,
\end{equation}
with $p$ the probability of initialising on the $A_1$ transition, and $\Gamma_\mathrm{A_1}\approx \SI{26}{\mega \hertz}$ and $\Gamma_\mathrm{A_2}\approx \SI{14}{\mega \hertz}$ the linewidths (FWHM) of the $A_1$ and $A_2$ transitions, respectively \cite{liu_silicon_2024}. For these values, Eq. \ref{eq:linewidth_weighted_mean} approximates the true linewidth with deviation less than $5\%$ (Supplementary Fig. \ref{fig:mean_linewidth}b).

A more rigorous treatment for estimating the individual $A_1$ and $A_2$ linewidths consists of explicitly specifying both transitions (separated by $\Delta \approx\SI{1}{\giga\hertz}$) with their respective amplitudes and lifetimes in the spectral response function $\lambda(f)$ (instead of using Eqs. \ref{eq:lorentzian_spectral_response} and \ref{eq:lzs_solution}). A more straightforward solution is to use an additional third laser (not available in this work) to perform the `check' block as is done in Fig. \ref{fig:spectral_diffusion_dynamics}a, which leaves no ambiguity for the initialisation configuration. Both the $A_1$ or $A_2$ transitions can then be measured individually by using a single NIR laser (together with the MW radiation) to probe around the two laser frequencies used in the `check' block. 

\clearpage
\subsection{Spectral diffusion dynamics during `scanning' PLE}
Complementary to the measurements in Fig. \ref{fig:lifetime_limited}, we perform repetitive PLE linescans of a V2 center (as in Refs. \cite{heiler_spectral_2024,babin_fabrication_2022,anderson_electrical_2019, orphal-kobin_optically_2023, koch_limits_2023}, among others). This is a different V2 center than the one studied in the main text, in particular, it resides in a different sample diced from the same wafer, but electron-irradiated at a slight lower dose of $\SI{2e12}{\per\centi\meter\squared}$. For fair comparison, we include the diffusion-averaged PLE measurement, and the NIR-lasers diffusion measurement (see Fig. \ref{fig:intro}f and as in Fig. \ref{fig:spectral_diffusion_dynamics}d of main text) in Supplementary Fig. \ref{fig:SI_Fast_PLE_plot}. These measurements indicate large diffusion-averaged linewidth and a high spectral diffusion rate (\SI{8(2)}{\giga\hertz/\second}). As noted in earlier work \cite{orphal-kobin_optically_2023,koch_limits_2023,heiler_spectral_2024}, summing individual scans may result in underestimation of the homogeneous linewidth due to limited photon statistics \cite{orphal-kobin_optically_2023}, or to overestimation of the linewidth if the emitter frequency changes during, and in between, repetitions. From the spectral diffusion measurements we can roughly distill the diffusion timescale to be $\sim \SI{10}{\milli \second}$. Since a single PLE linescan cannot be performed (with our current hardware) faster than the diffusion timescale, we limit the time during which the NIR laser is on. We scan the frequency by ramping an external voltage to the cavity of the laser (\SI{1.4}{\volt}). Each PLE scan takes \SI{500}{\milli \second}, where each voltage step (\SI{19}{\milli \volt}) the NIR-laser is turned on for \SI{1}{\milli \second} with \SI{10}{\nano \watt} of power, resulting in a total NIR-laser time of \SI{71}{\milli \second} during a single repetition.
\begin{figure}[ht]
  \includegraphics[width=1 \columnwidth]{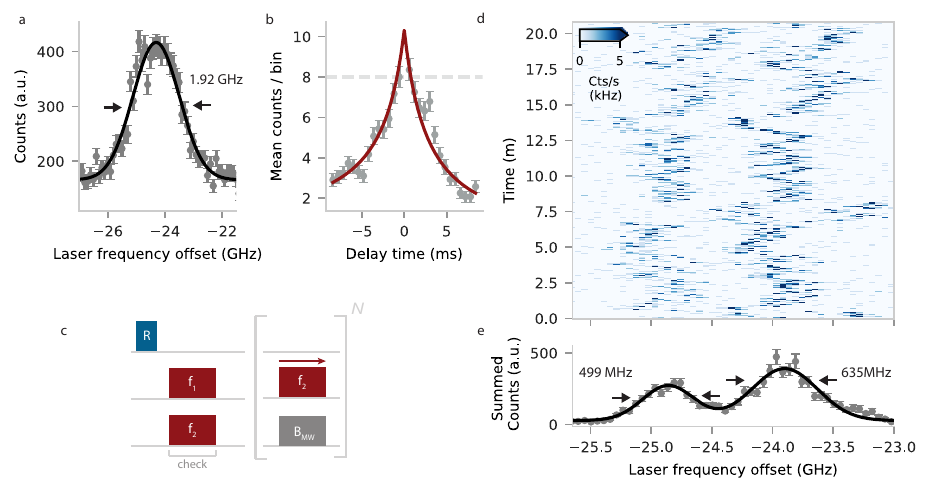}
  \caption{\label{fig:SI_Fast_PLE_plot} \textbf{Spectral diffusion during repetitive NIR-laser PLE scans.} \textbf{a)} Diffusion-averaged broadening of the defect in a different nanopillar (in a different sample) compared to the main text with $\sim\SI{1.92}{\giga \hertz}$ FWHM. \textbf{b)} Spectral diffusion measurement and fit under 10 nW of NIR laser excitation (916 nm) (same measurement as Fig. \ref{fig:spectral_diffusion_dynamics}). \textbf{c)} Experimental sequence used for the repetitive-scanning experiment. A `check' block is executed until the collected counts exceed a set threshold (here T = 30). Subsequently, the $f_2$ laser is scanned by ramping an external voltage to the cavity of the laser (\SI{1.4}{\volt}), while applying microwave resonant with the ground-state spin transition (\SI{6}{\volt}, different wire-sample distance compared to main text) with $N=500$ scans. Each scan takes \SI{500}{\milli \second}, where each voltage step (\SI{19}{\milli \volt}) the resonant laser is turned on for \SI{1}{\milli \second} with \SI{10}{\nano \watt} of power, resulting in a total resonant laser time of \SI{71}{\milli \second} during a single repetition. Before and after the voltage scan, the laser frequency is measured (\SI{1}{\second}) and a linear extrapolation between the start end and frequency is created to convert the applied voltage to an applied laser frequency. \textbf{d)} 500 PLE scans of the same defect as in b). Each repetition, the laser is scanned over $\sim\SI{3}{\giga \hertz}$). \textbf{e)} Summed counts of all the repetitions, indicating the A1 and A2 transition at \SI{-24.86}{\giga \hertz} and \SI{-23.90}{\giga \hertz} respectively, with FWHMs of \SI{499(14)}{\mega \hertz}, and \SI{635(11)}{\mega \hertz}, respectively (Gaussian fit). The laser frequency is offset to \SI{327.112}{\tera \hertz}.
}
\end{figure}

\clearpage

\subsection{Confocal microscopy setup}
In Supplementary Fig. \ref{fig:SI_Optical_Setup} the schematics of the optical setup is shown, which is divided into two parts, in-fiber (left) and free-space (right). The electronics are not depicted in the figure.\\
\textbf{In Fiber:} Two NIR lasers (\SI{916}{\nano \meter}, Toptica DL Pro and the Spectra-Physics Velocity TLB-6718-P, are frequency-locked to a wavemeter (HF-Angstrom WS/U-10U), using a 99:1 beamsplitter. Their optical power is modulated by acousto-optic-modulators (AOM, G\&H SF05958). The power of the \SI{785}{\nano \meter} repump laser (Cobolt 06-MLD785) is directly controlled via analog modulation. A wavelength division multiplexer (WDM, OZ Optics) combines the \SI{785}{\nano \meter} repump and \SI{916}{\nano \meter} NIR laser light, after which the light is coupled out to free space using a zoom fiber collimator (Thorlabs, ZC618APC). \\
\textbf{Free Space:} The collimated beam passes through a variable neutral density filter (ND, Thorlabs NDC-50C-4-B), after which a shortpass filter (Semrock, FF01-945/SP-25) is used to remove any residual noise from the NIR lasers. A $\frac{\lambda}{2}-\frac{\lambda}{4}$ waveplate combination allows for polarisation control. The excitation path and detection path are separated with a broadband 90:10 beamsplitter (Thorlabs, BS041). 

A flip mirror and 50:50 pellicle beamsplitter (Thorlabs, BP150) enable imaging of the sample with a visible LED (MCWHL6-C2) and a CCD-camera (ClearView Imaging, BFS-U3-16S2M-CS).

A 0.9 NA microscope objective (Olympus, MPLFLN 100x) is used to focus excitation light onto the nanopillars, and to collect fluorescence. The objective is kept at room temperature and under vacuum, and can be moved using a configuration of 3 piezo-electric stages (PI Q545.140). The sample is cooled down to \SI{4}{\kelvin} in a cryostat (Montana Instruments S100), while a heat shield kept at \SI{30}{\kelvin} limits thermal radiation from the objective. 

Collected fluorescence passes through a 90:10 beamsplitter, after which it can be routed either to a spectrometer (Princeton Instruments IsoPlane 81), filtered with an \SI{830}{\nano \meter} long-pass filter (Semrock, BLP01-830R-25), or to an avalanche photon detector (APD, Laser components, COUNT-50N) with an expected detection efficiency of 35\% at \SI{920}{\nano \meter} and 18\% at \SI{1000}{\nano \meter}. Next to a \SI{830}{\nano \meter} long pass filter, an additional long-pass filter (Semrock, FF01-937/LP-25), placed at an angle, is used to filter out reflected light originating from the NIR lasers (916 nm).\\
\textbf{Electronics}
Microwave pulses are generated with an arbitrary waveform generator (Zurich Instruments, HDAWG8), and subsequently amplified (Mini-circuits LZY-22+). A bondwire is spanned across the sample to deliver the MW radiation close to the sample surface ($\sim \SI{50}{\micro \meter}$). The coarse time scheduling ($\SI{1}{\micro \second}$ resolution) of the experiments is managed by a microcontroller (ADwin Pro II).

\begin{figure}[ht]
  \includegraphics[width=1 \columnwidth]{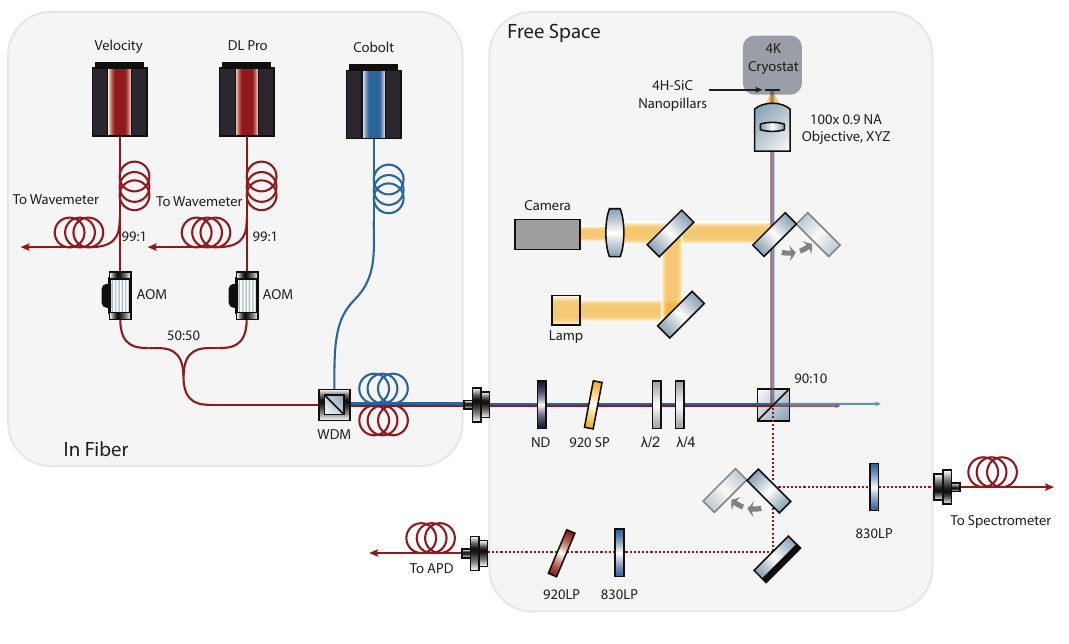}
  \caption{\label{fig:SI_Optical_Setup} \textbf{Optical setup:} Description of elements is given in the text. 
}
\end{figure}
\subsection{Literature review on silicon carbide quantum emitter linewidths}
\label{sec:PLE_literature_review}
In this section we provide an overview of the work done using silicon vacancies and divacancies in silicon carbide in the context of optical coherence. In table \ref{tab:your-table-label}, we report on multiple parameters that can impact the optical coherence. These parameters include the defect generation method, the annealing strategy, the measured optical absorption linewidth, material parameters of the layer that contains the investigated defect, and the use of a potential nanostructure for efficient light extraction.

\begin{sidewaystable}
    \centering
    \resizebox{\textwidth}{!}{%
    \begin{tabular}{lllllll} 
        \toprule 
        Work & Defect & Defect generation & Annealing & PLE linewidth & Sample parameters in the defect layer & Nanostructure\\
        \midrule 
        Banks 2019\cite{banks_resonant_2019} & V2 & e: 1e12 $\SI{}{\per\centi\meter\squared}$, 2 MeV & - & $\sim$65 MHz & 18$\SI{}{\micro\meter}$ n-doped epi layer. [N]=$\sim$3e14 $\SI{}{\per\centi\meter\cubed}$ & - \\
        
        Nagy 2019 \cite{nagy_highfidelity_2019} & V1 & e: 1e12 $\SI{}{\per\centi\meter\squared}$, 2 MeV & 30 m, 600\degree C & $\sim$60 MHz & 110$\SI{}{\micro\meter}$ [$^{\mathrm{28}}$Si]>99.85\%, [$^{\mathrm{13}}$C]>99.98\% epi layer. [N] = $\sim$3.5e13 $\SI{}{\per\centi\meter\cubed}$ & SIL \\
        
        Udvarhelyi 2020 \cite{udvarhelyi_vibronic_2020} & V1 \& V2 & e: 1e13 $\SI{}{\per\centi\meter\squared}$, 2 MeV & 30 m, 600\degree C & $\sim$100 MHz & 110$\SI{}{\micro\meter}$ [$^{\mathrm{28}}$Si]>99.85\%, [$^{\mathrm{13}}$C]>99.98\% epi layer. [N]=$\sim$3.5e13 $\SI{}{\per\centi\meter\cubed}$ & SIL \\

        Nagy 2021 \cite{nagy_narrow_2021} & V2 & e: 1e13 $\SI{}{\per\centi\meter\squared}$, 2 MeV & - & $\sim$300 MHz$^*$ & [N]=$\sim$5e13 $\SI{}{\per\centi\meter\cubed}$ & - \\

        Babin 2022 \cite{babin_fabrication_2022} & V2 & He$^{+}$: 1e11 $\SI{}{\per\centi\meter\squared}$, 6 keV & 30 m, 600\degree C & $\sim$25-40 MHz & 110$\SI{}{\micro\meter}$ [$^{\mathrm{28}}$Si]>99.85\%, [$^{\mathrm{13}}$C]>99.98\% epi layer. [N]=$\sim$4e13 $\SI{}{\per\centi\meter\cubed}$ & - \\
        
        Babin 2022 \cite{babin_fabrication_2022} & V2 & e: 5e11 $\SI{}{\per\centi\meter\squared}$, 2 MeV & 30 m, 600\degree C & Best: $\sim$30 MHz & 28$\SI{}{\micro\meter}$ [$^{\mathrm{28}}$Si]>99.85\%, [$^{\mathrm{13}}$C]>99.98\% epi layer. [N]=$\sim$3e15 $\SI{}{\per\centi\meter\cubed}$ & Triangular waveguides \\

        Lukin 2020 \cite{lukin_spectrally_2020} & V2 & e: 1e13 $\SI{}{\per\centi\meter\squared}$, 2 MeV & 30 m, 300\degree C & $\sim$100 MHz & 100$\SI{}{\micro\meter}$ $^{\mathrm{28}}$Si$^{\mathrm{13}}$C epi layer.& - \\

        Lukin 2020 \cite{lukin_spectrally_2020} & V2 & e: 5e12 $\SI{}{\per\centi\meter\squared}$, 23 MeV & 30 m, 300\degree C & $\sim$100 MHz & 100$\SI{}{\micro\meter}$ $^{\mathrm{28}}$Si$^{\mathrm{13}}$C epi layer.& - \\

        Lukin 2023 \cite{lukin_twoemitter_2023} & V2 & e: 1e13 $\SI{}{\per\centi\meter\squared}$, 2 MeV & 2 h, 550\degree C$^\dagger$ & $\sim$40 MHz & 20$\SI{}{\micro\meter}$ n-doped epi layer. [N]=$\sim$2e13 $\SI{}{\per\centi\meter\cubed}$ & Disk resonator \\

        Fang 2023 \cite{fang_experimental_2024} & V1 & e: 2.3e12 $\SI{}{\per\centi\meter\squared}$, 10 MeV & 30 m, 500\degree C & $\sim$60 MHz & 80$\SI{}{\micro\meter}$ epi layer. [N]=$\sim$2e13 $\SI{}{\per\centi\meter\cubed}$ & SIL \\

        Christle 2017 \cite{christle_isolated_2017} & VV & e: 5e12 $\SI{}{\per\centi\meter\squared}$, 2.5 MeV & 30 m, 750\degree C & $\sim$2 GHz & $\SI{730}{\micro\meter}$ sublimation epitaxially grown 3C-SiC layer. [N]$\sim$5e15 $\SI{}{\per\centi\meter\cubed}$ & - \\

        Christle 2017 \cite{christle_isolated_2017} & VV & e: 5e12 $\SI{}{\per\centi\meter\squared}$, 2 MeV & 30 m, 745\degree C & $\sim$100 MHz & $\SI{120}{\micro\meter}$ 4H-SiC epi layer. [N]<5e13 $\SI{}{\per\centi\meter\cubed}$ & - \\

        Miao 2019 \cite{miao_electrically_2019} & VV & e: 3e12 $\SI{}{\per\centi\meter\squared}$, 2 MeV & 30 m, 850\degree C & $\sim$21 MHz & 20$\SI{}{\micro\meter}$ i-type epi layer. [N]<1e15 $\SI{}{\per\centi\meter\cubed}$ & - \\

        Anderson 2019 \cite{anderson_electrical_2019} & VV & - & - & $\sim$31 MHz & - & - \\

        Crook 2020 \cite{crook_purcell_2020} & VV & e: 1e16 $\SI{}{\per\centi\meter\squared}$, 2 MeV & 30 m, 850\degree C & $\sim$4 GHz & n-p-n-i-n epi layers. ([N$_\mathrm{n}$]=1e18, [Al$_\mathrm{p}$]=1e18, [N$_\mathrm{i}$]<1e15) $\SI{}{\per\centi\meter\cubed}$ & Photonic crystal cavity \\

        Anderson 2022 \cite{anderson_fivesecond_2022} & VV & e: 1e13 $\SI{}{\per\centi\meter\squared}$, 2 MeV & 810\degree C & <0.5 GHz & 90$\SI{}{\micro\meter}$ [$^{\mathrm{28}}$Si]>99.85\%, [$^{\mathrm{13}}$C]>99.98\% epi layer. [N]=$\sim$6e13 $\SI{}{\per\centi\meter\cubed}$ & No nanostructure \\
        
        \bottomrule 
    \end{tabular}%
    }
    \caption{Sample parameters on literature publications that demonstrated PLE measurements with silicon vacancy (V1, V2) defects and divacancy (VV) defects in silicon carbide. A dash (-) indicates the information is not specified in the respective publication. $^*$: Increased linewidth attributed to global sample properties, such as an increased Fermi level due to the elevated [N], rather than defect density. $^\dagger$: Annealing tailored to HSQ bonding for sample preparation.}
    \label{tab:your-table-label}
\end{sidewaystable}

\end{document}